\documentclass[11pt,a4paper,floats]{article}

\usepackage{graphics}
\usepackage{epsf}
\usepackage{eufrak}
\usepackage{epsfig}

\def\rhoa{\rho_{\scriptscriptstyle \mathrm{a}}}

\begin{document} 

\flushbottom

\title{\flushleft \textbf Tricritical directed percolation}


\date{}
\maketitle

\setcounter{page}{1}
\markright{Journal of Statistical Physics {\textbf 123}, 193 (2006) }
\thispagestyle{myheadings}
\pagestyle{myheadings}

\noindent{\textbf {S.~L\"ubeck\footnote{Theoretische Physik,
Universit\"at Duisburg-Essen, 
47048 Duisburg, Germany}
}}\\[3mm]

\hfill\parbox{9.7cm}{
\footnotesize
{\textit {Received July 5, 2005; accepted September 30, 2005}}\\[-2mm]
\hrule
\vspace{1mm}
We consider a modification of the contact process 
incorporating higher-order reaction terms.
The original contact process 
exhibits a non-equilibrium phase transition 
belonging to the universality class of directed percolation.
The incorporated higher-order reaction 
terms lead to a non-trivial phase diagram.
In particular, a line of continuous phase transitions
is separated by a tricritical point from a line of
discontinuous phase transitions.
The corresponding tricritical scaling behavior 
is analyzed in detail, i.e., 
we determine the critical exponents, various 
universal scaling functions as well as universal amplitude
combinations. 
\\[-2mm]
\hrule
\vspace{2mm}
}

\vspace{10mm}




\section{\textsf{INTRODUCTION}}

The concept of universality is well established
for equilibrium critical phenomena where a 
unifying framework exists.
Compared to the equilibrium situation less 
is known in case of non-equilibrium phase transitions.
In particular a classification scheme of the rich
and often surprising variety of non-equilibrium phase 
transitions is still lacking.
Nevertheless, it is expected that analogous to 
equilibrium each non-equilibrium universality class
is characterized by a certain symmetry.
In this work we focus on stochastic processes which
exhibit irreversible phase transitions into absorbing
and fluctuation free states
(see~\cite{HINRICHSEN_1,ODOR_1,LUEB_35,MARRO_1} for recent reviews).
In that case the universality determining symmetry is expressed 
in a corresponding path integral formulation~\cite{ODOR_1,JANSSEN_4}
which is associated to the considered stochastic process.
A well known example is the universality class of 
directed percolation (DP).
According to its robustness and ubiquity  
(including critical phenomena
in physics, biology, as well as catalytic chemical reactions)
directed percolation is recognized as the paradigm 
of non-equilibrium phase transitions into absorbing states.
All non-equilibrium critical systems belong to the DP
universality class if the associated absorbing phase transition is 
described by a single component order parameter
and if the corresponding coarse grained system 
obeys the so-called rapidity reversal symmetry 
(at least asymptotically).
Unfortunately, this symmetry is usally reflected on a 
coarse grained level only.
Often, it is masked on a microscopic level, e.g.~it
is not reflected in the dynamic rules of certain lattice
models.
But the robustness of the DP universality class is
expressed by the conjecture of Janssen and 
Grassberger~\cite{JANSSEN_1,GRASSBERGER_2}:
short-range interacting systems, exhibiting a
continuous phase transition into an absorbing state,
belong to the DP universality class, if they are characterized
by a one-component order parameter and no additional symmetries
and no quenched disorder.
This robustness of the DP universality class
was recently demonstrated for five lattice models~\cite{LUEB_32}.
Despite the different interaction details
(such as different lattice structures,
different update schemes,
different implementation schemes of the conjugated field,
infinite and finite numbers of absorbing states, 
models with and without multiple particle occupation, etc.)
all considered models are characterized by the same
universal scaling functions.
Different universality classes than DP occur if the 
rapidity reversal is broken, 
e.g.~by quenched 
disorder~\cite{JANSSEN_11,JENSEN_10,MOREIRA_1,CAFIERO_1,HOOYBERGHS_1,VOJTA_2},
or additional symmetries such as 
particle-hole symmetry 
(compact directed percolation~\cite{DOMANY_1,ESSAM_1}),
particle conservation 
(Manna universality class~\cite{MANNA_2,ROSSI_1,LUEB_26}),
or parity conservation (for example branching annihilating
random walks with an even number of offsprings~\cite{ZHONG_1,CARDY_2}).

Another universality class of absorbing phase transitions
which is directly related to DP is tricritical directed 
percolation (TDP).
Analogous to the $\phi^6$-theory in equilibrium critical
phenomena the 
process of TDP incorporates higher-order reaction terms
than ordinary DP.
Investigations of TDP in one-dimensional systems 
traces back to the seminal work of 
Grassberger~\cite{GRASSBERGER_2}.
Later, a field theoretical analysis was performed by  
Ohtsuki and Keyes~\cite{OHTSUKI_1,OHTSUKI_2} 
(see also~\cite{JANSSEN_4,JANSSEN_14}).
In that work the authors discuss a one-component reaction-diffusion 
system that exhibits a tricritical point.
This tricritical point separates continuous DP-like transitions
from first-order transitions.
Using an $\epsilon$-expansion several 
critical exponents were estimated.
Furthermore, the authors determine  
the upper critical dimension~$D_{\scriptscriptstyle \mathrm{c}}=3$.
Compared to the established field theory of TDP
less work was done numerically.
Several modifications of one-dimensional DP-lattice models are known 
which yield multicritical behavior.
But mostly a bicritical point
is observed or the 
occurring tricriticality is analyzed within a mean field
level only (see e.g.~\cite{BASSLER_1,BAGNOLI_2,ATMAN_1}).
Surprisingly no systematic numerical investigations of 
tricritical-DP scaling behavior were performed so far.

In this work we analyze the tricritical contact process (TCP).
This model is a modification of the 
well-known contact process~\cite{HARRIS_2}
which belongs to the universality class of directed
percolation.
The modification incorporates higher-order particle reactions,
in particular a pair reaction scheme is added. 
It is worth mentioning that multicritical behavior
does not necessarily appear by introducing additional 
higher-order reaction schemes.
Multicritical behavior occurs if the resulting
lower-order reactions vanish on a coarse grained scale
(see~\cite{OHTSUKI_2} and references therein).
In our case, the added pair reaction scheme
leads to a non-trivial phase diagram 
containing first- and second-order transitions as well
as a tricritical point.
In $D\ge 2$ the tricritical point separates a line of
DP-like transitions from a line of first-order transitions.
In one-dimension,  
preliminary results do indicate that the TCP does not exhibit 
a tricritical behavior.
According to numerical simulations and to a mean field cluster
analysis~\cite{ODOR_PRIVAT_2005} 
the first-order line and the tricritical
point collapse for $D=1$ to the end point of the corresponding
phase diagram (this agrees with the more general argument
that first-order transitions can not occur in fluctuating one-dimensional
systems due to the fact that the surface tension of a given domain
does not depend on its size).
Therefore, we focus on the higher-dimensional systems in the following.
First, we survey the mean field behavior of TDP and of the TCP 
which are valid for $D>3$.
Second, the two-dimensional TCP is numerically examined.
A scaling analysis of the DP-like transitions and of the
tricritical phase transition itself is presented.
In particular, the full crossover between both universality
classes is recovered.
The obtained values of universal quantities, 
such as the critical exponents, are compared to the 
results of the corresponding field 
theories~\cite{JANSSEN_4,OHTSUKI_1,OHTSUKI_2}.

Furthermore, the regime of first-order transitions is 
investigated.
Discontinuous phase transitions are of their own 
interest (see for example~\cite{DICKMAN_18,BAGNOLI_2,ZIFF_2,BROSILOW_1})
because
most absorbing phase transitions are of second-order.
As well known, discontinuous transitions are usually 
characterized by hysteresis cycles caused by the coexistence
of two phases.
A small but finite hysteresis can be observed in numerical
simulations of the TCP.

\section{\textsf{TRICRITICAL DIRECTED PERCOLATION}}

The process of directed percolation might be 
represented by the Langevin
equation (see e.g.~\cite{JANSSEN_1}) 
\begin{equation}
\lambda^{-1}\, \partial_{\scriptscriptstyle t} \rhoa
= \tau  \rhoa -  {g}  \rhoa^2 - c \rhoa^3 
+ \Gamma \, \nabla^2  \rhoa
+ h + \eta \, .
\label{eq:langevin_dp_01}
\end{equation}
The particle density on a mesoscopic (coarse grained) scale
$\rhoa=\rhoa({\underline x},t)$ corresponds 
to the order parameter of the non-equilibrium phase transition.
The control parameter of the transition~$\tau$
describes the distance from the critical point~$\tau=0$.
A finite positive particle density occurs above the
transition point ($\tau>0$) whereas the absorbing vacuum state
($\rhoa=0$) is approached below the transition point.
The external field~$h$ is conjugated to the order parameter
and is usually implemented as a spontaneous particle creation
process~(see e.g.~\cite{LUEB_27}).
Furthermore, $\eta=\eta({\underline x},t)$ denotes the noise which 
accounts for fluctuations of the particle density.
The noise $\eta$ is a Gaussian random variable with zero mean
and whose correlator is given by~\cite{JANSSEN_1}
\begin{equation}
\langle \, \eta({\underline x},t)  
\eta({\underline x}^{\prime},t^{\prime})  \rangle
=  \lambda^{-1} \kappa  \, \rhoa ({\underline x},t) 
\, \delta({\underline x}-{\underline x}^{\prime}) 
\, \delta(t-t^{\prime})  .
\label{eq:langevin_dp_corr_01}
\end{equation}
Notice, that the noise ensures that the 
systems is trapped in the absorbing state \mbox{$\rhoa({\underline x},t)=0$}.
Furthermore, higher-order terms such as 
$\rhoa({\underline x},t)^3$,
$\rhoa({\underline x},t)^4, \ldots$ 
or 
$\nabla^4 \rhoa({\underline x},t),\nabla^6 
\rhoa({\underline x},t), \ldots$ are usually neglected
because they are irrelevant under renormalization group transformations
as long as $g > 0$.
Negative values of $g$ give rise to a first-order
phase transition whereas $g=0$ is associated with
a tricritical point~\cite{OHTSUKI_1,OHTSUKI_2}. 
In the latter cases the cubic term $\rhoa^3$ cannot be neglected.

In the following we present a simple but instructive mean
field treatment.
Neglecting the noise term as well as spatial variations
of the order parameter the steady state behavior 
($\partial_{\scriptscriptstyle t} \rhoa =0$) at zero field is given by
\begin{equation}
\rhoa=0 \quad \vee \quad
\rhoa=-\frac{g}{2c} \pm 
\sqrt{\frac{\tau}{c} + \left ( \frac{g}{2c}\right )^2\,} \, .
\label{eq:triDP_op_steady_state}
\end{equation}
The first solution corresponds to the absorbing phase.
According to a linear stability analysis it is
stable for $\tau<0$ and unstable for ($\tau>0$).
The solution with the $-$\,sign yields unphysical and unstable
results.
The $+$\,sign solution describes the order parameter as a function
of the control parameter~$\tau$ and of the additional scaling field~$g$.
For $g<0$, this solution is stable if $\tau>-g^2/4c$, otherwise 
it is unstable. 
Assuming that the system is in the active phase ($\rhoa>0$)
the order parameter jumps at the borderline ($\tau=-g^2/4c$)
from $\rhoa=|g|/2c$ to zero.
Thus the absorbing and the active phase coexist between the
spinodal $\tau=-g^2/4c$ and the line $\tau=0$ for $g<0$.

The tricritical behavior is obtained for $g=0$.
The order parameter behaves for $\tau>0$ as
\begin{equation}
\rhoa = \left ( {\tau}/{c} \right )^{\beta_{\scriptscriptstyle \mathrm{t}}}
\label{eq:triDP_op_rhoa_tau}
\end{equation}  
with the tricritical order parameter 
exponent~$\beta_{\scriptscriptstyle \mathrm{t}}=1/2$.
For $g>0$, the active phase is stable for $\tau>0$ and unstable
otherwise.
Close to the transition points ($\tau=0$)
the order parameter exhibits a 
directed percolation like behavior 
\begin{equation}
\rhoa= \left . -\frac{g}{2c} + 
\sqrt{\frac{\tau}{c} + \left ( \frac{g}{2c}\right )^2\,} 
\right |_{{\tau}\ll \frac{g^2}{4c}}
= \left ( \frac{\tau}{g} \right )^{\beta_{\scriptscriptstyle \mathrm{DP}}}
+ {\mathcal{O}}(\tau^2)
\label{eq:triDP_op_dp_like}
\end{equation}
with the exponent $\beta_{\scriptscriptstyle \mathrm{DP}}=1$.
The amplitude of this power-law diverges for $g\to 0 $, signaling the
crossover to the tricritical behavior.

\begin{figure}[t]
  \centering
  \includegraphics[width=10.0cm,angle=0]{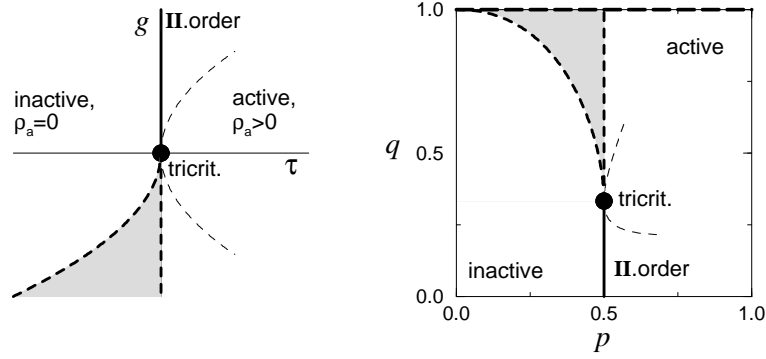}
  \caption{
    The mean field phase diagrams of tricritical directed
    percolation (TDP).
    The left figure sketches the phase diagram as a function
    of the coarse grained variables $\tau$ and $g$ 
    [see Eq.\,(\protect\ref{eq:langevin_dp_01})].
    The right figure shows the corresponding phase diagram 
    as a function of the microscopic variables~$p$ and~$q$,
    as defined by a reaction scheme of the tricritical contact 
    process (see text).
    The thick lines correspond to second-order phase
    transitions (II.order) with $\beta_{\scriptscriptstyle \mathrm{DP}}=1$. 
    The bold circles mark the transition point of 
    TDP with $\beta_{\scriptscriptstyle {\mathrm{t}}}=1/2$.
    The shadowed areas indicate the coexistence of the 
    absorbing and the active phase.
    The thin dashed lines illustrate the crossover to the tricritical
    behavior.
    In the right figure, two absorbing phases appear
    for $q=1$ (long dashed lines).
    These phases are the fully occupied lattice 
    (stable for $q>0$) and the empty lattice (stable for
    $q<1/2$).
    }
  \label{fig:phasedia_mf} 
\end{figure}

Sufficiently away from the critical line $\tau=0$ we observe
again the tricritical behavior 
\begin{equation}
\rhoa= \left . -\frac{g}{2c} + 
\sqrt{\frac{\tau}{c} + \left ( \frac{g}{2c}\right )^2\,} 
\right |_{{\tau}\gg \frac{g^2}{4c}}
= \sqrt{\frac{\tau}{c}} + {\mathcal{O}}(\tau^{-1})
\label{eq:triDP_op_tridp_like}
\end{equation}
for both $g>0$ and $g<0$.
The crossover from the tricritical behavior
takes place at  
\begin{equation}
{\tau} = {\mathcal{O}}(g^2/4c) \, .
\label{eq:triDP_crossover}
\end{equation}
The complete phase diagram is sketched in 
Fig.\,\ref{fig:phasedia_mf}.

Within the active phase for $g\ge 0$ the order parameter
obeys the scaling form
\begin{equation}
\rhoa = \lambda^{-{\beta_{\scriptscriptstyle \mathrm{t}}}} \;
{\tilde r}_{\scriptscriptstyle {\mathrm{tDP}}}
(\lambda \tau,\,  g \lambda^\phi,\,  h=0)
\label{eq:triDP_scaling_mf_01}
\end{equation}
for all positive $\lambda$ and 
with the crossover exponent $\phi=1/2$.
In particular, $\lambda=g^{-1/\phi}$ leads to the scaling form
\begin{equation}
\frac{\rhoa}{g^{\beta_{\scriptscriptstyle \mathrm{t}}/\phi}} 
=  {\tilde r}_{\scriptscriptstyle {\mathrm{tDP}}}
(\tau g^{-1/\phi}, 1 , 0)
\label{eq:triDP_scaling_mf_02}
\end{equation}
with 
\begin{equation}
{\tilde r}_{\scriptscriptstyle {\mathrm{tDP}}}(x, 1 , 0) \sim
\left \{
\begin{array}{ll}
x^\beta_{\scriptscriptstyle \mathrm{DP}}  & {\mathrm{if}} \quad x\ll 1\\[2mm]
x^{\beta_{\scriptscriptstyle \mathrm{t}}} & {\mathrm{if}} \quad x\gg 1 \, .
\end{array}
\right .
\label{eq:triDP_scaling_mf_03}
\end{equation}
In that way, the scaling plot 
$\rhoa/g^{\beta_{\scriptscriptstyle \mathrm{t}}/\phi}$
vs.~$\tau g^{-1/\phi}$ reflects the crossover from DP
to TDP.

Now we consider the scaling behavior at non-zero 
conjugated field~$h$.
The tricritical order parameter
behavior at criticality ($\tau=0$ and $g=0$) is given by
\begin{equation}
\rhoa = (h/c)^{\beta_{\scriptscriptstyle \mathrm{t}}/
\sigma_{\scriptscriptstyle \mathrm{t}}}
\label{eq:triDP_op_rhoa_h}
\end{equation}
with the exponent $\sigma_{\scriptscriptstyle \mathrm{t}}=3/2$.
At the tricritical point the equation of state obeys
the scaling form
\begin{equation}
\rhoa = \lambda^{-{\beta_{\scriptscriptstyle \mathrm{t}}}} \;
{\tilde r}_{\scriptscriptstyle \mathrm{tDP}}
(\lambda \tau,\,  g=0, 
\,  h \lambda^{\sigma_{\scriptscriptstyle \mathrm{t}}}) \, .
\label{eq:triDP_scaling_mf_field_01}
\end{equation}
Instead of the above equation the so-called Widom-Griffiths scaling 
form 
\begin{equation}
h = \lambda^{-{\sigma_{\scriptscriptstyle \mathrm{t}}}} \;
{\tilde h}_{\scriptscriptstyle {\mathrm{tDP}}}
(\lambda \tau,\,  \rhoa \lambda^{\beta_{\scriptscriptstyle \mathrm{t}}},
\,  g=0) 
\label{eq:triDP_scaling_mf_field_02}
\end{equation}
is often used to describe the scaling behavior of the
equation of state.
Despite metric factors the tricritical scaling form is
given by 
\begin{equation}
{\tilde h}{\scriptscriptstyle {\mathrm{tDP}}}(x,y,g=0) = -x y + y^3 \, .
\label{eq:triDP_scaling_mf_field_03}
\end{equation}
The corresponding scaling form of ordinary DP
${\tilde h}_{\scriptscriptstyle {\mathrm{DP}}}(x,y)=-x y + y^2$
reflects the different universality class.

For the sake of completeness we present the 
dynamical order parameter behavior at tricriticality
\begin{eqnarray}
\rhoa(\tau=0,g=0,h=0,t) & = &
\left (\rho_{\scriptscriptstyle \mathrm{a},t=0}^{-2}+2ct\, \right)^{-1/2}
\nonumber \\[2mm]
& \mathop{\longrightarrow}\limits_{t\to \infty}  &
(2ct)^{-\alpha_{\scriptscriptstyle \mathrm{t}}}
\label{eq:triDP_dyn_op_01}
\end{eqnarray}
with $\alpha_{\scriptscriptstyle \mathrm{t}}=1/2$.
Furthermore, the steady state value of the order parameter
[Eq.\,(\ref{eq:triDP_op_rhoa_tau})]
is approached from below as
\begin{equation}
\rhoa(\tau,g=0,h=0,t) = 
\sqrt{\frac{\tau}{c}\,} \, 
\left [ 1 - c_{\scriptscriptstyle 0} 
{\mathrm{e}}^{-t/\xi_{\scriptscriptstyle \parallel}}
+ {\mathcal{O}}({\mathrm{e}}^{-2 t/\xi_{\scriptscriptstyle \parallel}}) 
\right ] .
\label{eq:triDP_dyn_op_02}
\end{equation}
Here, the constant~$c_{\scriptscriptstyle 0}$ contains the initial 
conditions and $\xi_{\scriptscriptstyle \parallel}$ denotes
the temporal correlation length 
\begin{equation}
\xi_{\scriptscriptstyle \parallel} = 
(2 \tau)^{-\nu_{\scriptscriptstyle \parallel,{\mathrm{t}}}}
\label{eq:triDP_corr_time_mf}
\end{equation}
with $\nu_{\scriptscriptstyle \parallel,{\mathrm{t}}}=1$.
Incorporating spatial variations of the order parameter
the spatial correlation length~$\xi_{\scriptscriptstyle \perp}$ 
can be derived via an Ornstein-Zernicke approach.
The resulting correlation length exponent
$\nu_{\scriptscriptstyle \perp,{\mathrm{t}}}=1/2$
leads to the dynamical exponent
$z_{\scriptscriptstyle \mathrm{t}}=
\nu_{\scriptscriptstyle \parallel,{\mathrm{t}}}/
\nu_{\scriptscriptstyle \perp,{\mathrm{t}}}=2$.
Eventually we just mention that the tricritical exponent
of the order parameter fluctuations is given by
$\gamma_{\scriptscriptstyle \perp,{\mathrm{t}}}^{\prime}=1/2$~\cite{OHTSUKI_1}.
This value reflects a qualitative difference between the
mean field scaling behavior of DP and TDP.
In the latter case the fluctuations diverge at the
transition point whereas they remain finite
(jump) in case of DP~\cite{MORI_1,LUEB_28}.
Another difference between both universality classes concerns
the value of the upper critical dimension, 
namely $D_{_{\scriptscriptstyle {\mathrm{c}}}}=4$ for 
ordinary DP~\cite{CARDY_1,OBUKHOV_2}
and $D_{_{\scriptscriptstyle {\mathrm{c}}}}=3$ 
for TDP~\cite{OHTSUKI_1,OHTSUKI_2}.


\section{\textsf{MODEL AND SIMULATIONS}}
\label{sec:model}

We consider a modification
of the contact process (CP).
The CP was introduced by Harris in order to
model the spreading of epidemics~\cite{HARRIS_2}.
It is a continuous-time Markov
process that is usually defined on a $D$-dimensional
simple cubic lattice.
A lattice site may be empty ($n=0$) 
or occupied ($n=1$) by a particle and the 
dynamics is characterized by spontaneously occurring
processes, taking place with certain transition rates.
In numerical simulations the asynchronous 
update is realized by a random sequential update 
scheme:~A particle on a randomly selected lattice site~${i}$ 
is annihilated with rate one, 
whereas particle creation takes places on an 
empty neighboring site with rate $\lambda N / 2 D$, i.e.,
\begin{eqnarray}
\label{eq:trans_rate_cp_annih}
n_{\scriptscriptstyle i} = 1 \quad  
&  \mathop{\longrightarrow}\limits_{1} &
\quad n_{\scriptscriptstyle i} = 0 \, ,\\
\label{eq:trans_rate_cp_creat}
n_{\scriptscriptstyle i} = 0 \quad  
&  \mathop{\longrightarrow}\limits_{\lambda N / 2 D} & 
\quad n_{\scriptscriptstyle i} = 1 \, ,
\end{eqnarray}
where~$N$ denotes the number of occupied neighbors
of~$n_{\scriptscriptstyle i}$.
Note that the rates are defined as transition probabilities
per time unit, i.e., they may be larger than one.
Thus, rescaling the time will change the transition rates.
In simulations a discrete time formulation of the
contact process is performed.
In that case a particle creation takes place at a randomly 
chosen neighbor site with probability $p=\lambda/(1+\lambda)$
whereas particle annihilation occurs with 
probability $1-p=1/(1+\lambda)$.


In the language of reaction-diffusion models the 
contact process is described by the reduced reaction scheme
\begin{equation}
A \longrightarrow 0 \, , \quad
A \longrightarrow 2 A \,
\label{eq:reac_scheme_cp}
\end{equation}
which is controlled by the parameter~$p$ and 
where the quantity~$A$ represents a particle.
According to Ohtsuki and Keyes~\cite{OHTSUKI_1,OHTSUKI_2}
higher-order reactions may lead to a tricritical behavior as well
as to a first-order behavior.
Here, we additionally take into consideration
the pair reaction 
\begin{equation}
2A \longrightarrow 3A  \, .
\label{eq:reac_scheme_cp_pair}
\end{equation}
This reaction is controlled by a parameter~$q$.
Updating a given occupied lattice site~$i$, we first
perform the pair reaction scheme with probability~$q$
(otherwise a usual CP-update step is performed).
If a randomly selected neighbor of~$i$ is occupied
we add a third particle at an empty (also randomly selected)
neighbor of the pair.
If the lattice site~$i$ is isolated a
usual update procedure of the CP is performed.
The detailed reaction scheme is listed in Table\,\ref{table:reactions}.

It is essential for the following that 
within our implementation the annihilation processes
\begin{equation}
A \longrightarrow 0 \, , \quad
AA \longrightarrow 0A \, , \quad
AA \longrightarrow A0 \,
\label{eq:reac_scheme_anni_01}
\end{equation}
take place with probabilities proportional to $(1-q)$.
Thus these reactions are suppressed for $q\approx 1$
and are eventually forbidden for $q=1$.
In that case particle annihilation occur via the reactions
\begin{equation}
A0 \longrightarrow 00 \, , \quad
0A \longrightarrow 00 \, .
\label{eq:reac_scheme_anni_02}
\end{equation}
This implies that annihilation processes are restricted
to the domain boundaries and do not occur insight of a domain.
This behavior is reminiscent of compact directed
percolation~\cite{DOMANY_1,LIGGETT_1,ESSAM_1} where the density of 
particles exhibits a discontinuous behavior at the critical
point.

\begin{table}[b]
\centering
\caption{
The reaction schemes and the corresponding probabilities 
of the considered tricritical contact process.
For $q=0$ the reaction scheme of the original
contact process is recovered (see text).
}
\vspace{0.3cm}
\label{table:reactions}
\begin{tabular}{|c|c|}
\hline
reaction &  probability\\  
\hline 
$A \longrightarrow 0$ 		& $\;(1-q)(1-p)\;$ \\
$0A \longrightarrow 00$ 	& $\;q(1-p)\;$ \\
$0A \longrightarrow AA$ 	& $\;(1-q)p\;$ \\
$AA \longrightarrow 0A$ 	& $\;(1-q)(1-p)\;$ \\
$0AA \longrightarrow AAA$ 	& $\;q\;$ \\
\hline
\end{tabular}
\end{table}

The above scenario is also reflected by a corresponding
mean field analysis.
Within the simplest mean field approach (single site approximation), 
the order parameter is given by
\begin{equation}
\partial_{\scriptscriptstyle t} \rhoa
= \tau_{\scriptscriptstyle p,q}  \rhoa -  
g_{\scriptscriptstyle p,q}  \rhoa^2 - 
c_{\scriptscriptstyle p,q} \rhoa^3 \,
\label{eq:langevin_tridp_01}
\end{equation}
with 
\begin{equation}
\tau_{\scriptscriptstyle p,q}=2p-1\, , \quad
g_{\scriptscriptstyle p,q}= p -(2-p) q  \, , \quad 
c_{\scriptscriptstyle p,q}=q \, .
\label{eq:coarse_grained}
\end{equation}
The tricritical point 
$(p_{\scriptscriptstyle \mathrm{t}}=1/2,q_{\scriptscriptstyle \mathrm{t}}=1/3)$ 
is determined by the conditions
$\tau=0$ and $g=0$.
The line of second-order phase transitions 
($p_{\scriptscriptstyle \mathrm{c}}=1/2$ 
and $q<1/3$) 
is obtained
from $\tau=0$ and $g>0$.
Within the first-order regime ($g<0$)
the absorbing phase is stable for $p<1/2$.
The borderline of stability of the 
active phase ($\rhoa>0$) is 
determined by $\tau=-g^2/4c$.
The corresponding mean field phase diagram is sketched in
Fig.\,\ref{fig:phasedia_mf}. 

For $q=1$ the corresponding mean field differential 
equation is given by
\begin{equation}
\partial_{\scriptscriptstyle t} \rhoa
= (2p-1)  \rhoa (1-\rhoa) +
\rhoa^2 (1-\rhoa) \, .
\label{eq:langevin_tcp_qe1}
\end{equation}
Obviously, the steady state solutions are the 
empty lattice ($\rhoa=0$ stable for $p<1/2$) and the
fully occupied lattice ($\rhoa=1$ stable for $p>0$).
Note that both phases coexist for $p<1/2$.
It is worth to compare the TCP for $q=1$ to the
process of compact directed percolation which 
is described by the equation (see e.g.~\cite{JANSSEN_4})
\begin{equation}
\partial_{\scriptscriptstyle t} \rhoa
= (2p-1)  \rhoa (1-\rhoa) \, .
\label{eq:langevin_cdp_01}
\end{equation}
Here, the fully occupied lattice is stable for $p>1/2$
whereas the empty lattice is stable for $p<1/2$.
Thus the process of compact directed percolation displays
no phase coexistence in contrast to the 
tricritical contact process for~$q=1$.
Furthermore, Eq.\,(\ref{eq:langevin_cdp_01}) clearly
express the particle-hole symmetry which is
the characteristic symmetry of the universality class
of compact directed percolation.
In case of the TCP for $q=1$, the particle-hole symmetry
is broken by the pair reaction processes contributing
$\rhoa^2(1-\rhoa)$ to the equation of motion.

Analyzing numerically or experimentally
the scaling behavior of tricritical systems
it is crucial to determine the value of the tricritical
point with high accuracy.
Thus a sensitive criterion is required to distinguish a 
first-order transition from a second-order transition.
At first glance, one is tempted to make use of the 
order parameter jump at the first-order transition.
At the tricritical point the order parameter changes
its behavior from a discontinuous jump to a continuous
power-law.
But first, this behavior is affected close to the
tricritical point by crossover effects
(see Fig.\,\ref{fig:phasedia_mf}).
Second, it is notoriously difficult to
distinguish a continuous phase transition with 
a small but finite value of the 
exponent~$\beta_{\scriptscriptstyle \mathrm{t}}$ 
from a discontinuous jump.
Especially this situation occurs in 
case of two-dimensional TDP.
An alternative way is to investigate instead of the
order parameter the order parameter fluctuations
which diverge at the tricritical point but remain
finite within the first-order regime.
But the fluctuation measurements suffer by crossover effects
in a similar way as the order parameter, i.e.,
a corresponding analysis yields therefore no 
significant improvement.

In order to circumvent these problems we apply a method
of analyzing that is based on the scaling behavior 
of the order parameter close to the tricritical point 
within the second-order transition regime.
In particular, the scaling form of the
order parameter is used to recover the complete crossover from 
ordinary DP to the TDP.
We assume that the order parameter as well as the
order parameter fluctuations obey the scaling forms
\begin{eqnarray}
\label{eq:triDP_scaling_2d_op}
\rhoa \sim \lambda^{-{\beta_{\scriptscriptstyle \mathrm{t}}}} \;
{\tilde r}_{\scriptscriptstyle {\mathrm{tDP}}}
(\lambda \tau,\,  g \lambda^\phi,\,  h=0)\\[2mm]
\label{eq:triDP_scaling_2d_fl}
\Delta\rhoa \sim \lambda^{{\gamma^{\prime}_{\scriptscriptstyle \mathrm{t}}}} \;
{\tilde d}_{\scriptscriptstyle {\mathrm{tDP}}}
(\lambda \tau,\, g \lambda^\phi,\,  h=0)
\end{eqnarray}
with the so far unknown tricritical exponents 
$\beta_{\scriptscriptstyle \mathrm{t}}$, 
$\gamma^{\prime}_{\scriptscriptstyle \mathrm{t}}$, and
$\phi$.
Asymptotically, the 
scaling functions have to fulfill the power-laws
\begin{eqnarray}
\label{eq:triDP_scaling_2d_op_asym}
{\tilde r}_{\scriptscriptstyle {\mathrm{tDP}}}(x, 1, 0) & \sim &
\left \{
\begin{array}{ll}
x^{\beta_{\scriptscriptstyle \mathrm{DP}}}   & {\mathrm{if}} \quad x\ll 1\\[2mm]
x^{\beta_{\scriptscriptstyle \mathrm{t}}} & {\mathrm{if}} \quad x\gg 1 \, ,
\end{array}
\right . \\[2mm]
\label{eq:triDP_scaling_2d_fl_asym}
{\tilde d}_{\scriptscriptstyle {\mathrm{tDP}}}(x, 1, 0) & \sim &
\left \{
\begin{array}{ll}
x^{-\gamma^{\prime}_{\scriptscriptstyle \mathrm{DP}}}   
& {\mathrm{if}} \quad x\ll 1\\[2mm]
x^{-\gamma^{\prime}_{\scriptscriptstyle \mathrm{t}}} 
& {\mathrm{if}} \quad x\gg 1 \, .
\end{array}
\right .
\end{eqnarray}
The values of the exponents~$\beta_{\scriptscriptstyle {\mathrm{DP}}}$
and~$\gamma^{\prime}_{\scriptscriptstyle {\mathrm{DP}}}$
are known with sufficient accuracy 
(see Table\,\ref{table:exponents} and references therein).
A serious problem is caused by the fact that the 
scaling forms contain the coarse grained variables $\tau$
and $g$.
In general, these variables depend on the (microscopic) 
parameters $p$ and $q$ in an unknown way.
But there exists a certain window of scaling
where the coarse grained variables can be replaced by the
model parameters $p$ and $q$.
This will allow the
determination of the tricritical point as well as of the 
exponents~$\beta_{\scriptscriptstyle {\mathrm{t}}}$
and $\gamma^{\prime}_{\scriptscriptstyle {\mathrm{t}}}$.

In the following the system is investigated within the active phase
close to the line of second-order phase transitions, 
i.e., $g(p,q)>0$ and $\tau(p,q)$ close to zero.
Performing simulations, the order parameter is measured
as a function of~$p$ for fixed~$q$.
Sufficiently close to the transition line $p_{\scriptscriptstyle \mathrm{c}}(q)$
the coarse grained variable is approximated
\begin{eqnarray}
\label{eq:tau_at_qc_01}
\tau(p,q) & = & \tau(p_{\scriptscriptstyle \mathrm{c}}(q)+\delta p, q) 
\nonumber \\[2mm]
& \approx & 
\underbrace{\tau(p_{\scriptscriptstyle \mathrm{c}}(q), q)}_{=0} \, + \, 
\left . \frac{\partial \tau}{\partial p} 
\right |_{p_{\scriptscriptstyle \mathrm{c}}(q)} \delta p 
\end{eqnarray}
with $\delta p = p - p_{\scriptscriptstyle \mathrm{c}}(q)$.
Since the derivative is taken at the transition line 
$p_{\scriptscriptstyle \mathrm{c}}(q)$ its value depends
on the parameter~$q$.
But close to the tricritical point we yield up to higher-orders
\begin{equation}
\tau(p,q) \;  \approx \; 
\left . \frac{\partial \tau}{\partial p} 
\right |_{p_{\scriptscriptstyle \mathrm{c}}
(q_{\scriptscriptstyle \mathrm{t}})} \delta p 
\, + \, {\mathcal{O}}(\delta p \, \delta q)
\label{eq:tau_at_qc_02}
\end{equation}
with $\delta q = q_{{\scriptscriptstyle \mathrm{t}}}-q$.

\begin{figure}[t]
  \centering
  \includegraphics[width=7.0cm,angle=0]{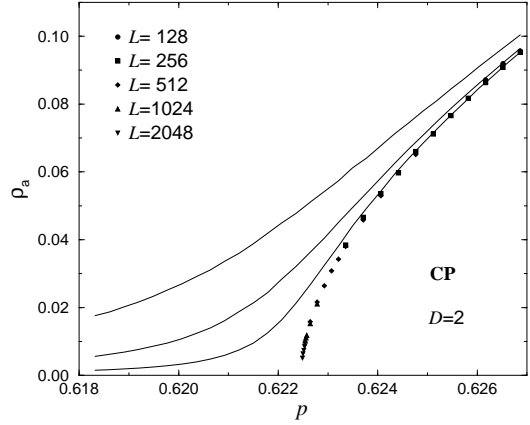}
  \caption{
    The order parameter~$\rhoa$ of the two-dimensional contact
    process (CP) as a function of the control parameter~$p$.
    The data are obtained from simulations on a simple cubic lattices
    of linear size~$L$ 
    for various field values (from $10^{-6}$ to $h=10^{-5}$).
    For non-zero field~$\rhoa$ exhibits an analytical behavior (lines).
    For zero field (symbols) the order parameter 
    vanishes at the transition point $p=0.62246$~\cite{DICKMAN_7}.
    }
  \label{fig:cp_rho_a_01} 
\end{figure}

A similar approximation can be obtained for $g(p,q)$.
In the vicinity of the critical line we use the
approximation
\begin{eqnarray}
\label{eq:q_at_qc_01}
g(p,q) & = & g(p_{\scriptscriptstyle \mathrm{c}}(q)+\delta p, q) 
\nonumber \\[2mm]
& \approx & 
g(p_{\scriptscriptstyle \mathrm{c}}(q), q) \, + \, 
\left . \frac{\partial g}{\partial p} 
\right |_{p_{\scriptscriptstyle \mathrm{c}}(q)} \delta p \, .
\end{eqnarray}
Sufficiently close to the tricritical point we obtain
\begin{eqnarray}
\label{eq:q_at_qc_02}
g(p,q) &   \approx  & 
\left .
\nabla g \cdot 
{\underline t}^{\vphantom X} \, 
\right |_{
p_{\scriptscriptstyle \mathrm{t}},q_{\scriptscriptstyle \mathrm{t}}}
\delta q \nonumber \\[2mm]
& + & 
\left .
\frac{\partial g}{\partial p} 
\right |_{
p_{\scriptscriptstyle \mathrm{t}},q_{\scriptscriptstyle \mathrm{t}}}
\, \delta p
\, + \, \mathcal{O}(\delta p \, \delta q) \, ,
\end{eqnarray} 
with $\nabla g= (\partial_p g, \partial_q g)$ and where
the tangential vector along the phase boundary is 
denoted by 
${\underline t}=(\partial 
p_{\scriptscriptstyle \mathrm{c}}(q)/\partial q, 1)$.
Assuming that $\nabla g \cdot {\underline t}$
and $\partial g / \partial p$ are of the same order
at the tricritical point,
the coarse grained variable $g(p,q)$ can be 
replaced in the scaling functions 
Eqs.\,(\ref{eq:triDP_scaling_2d_op_asym},
\ref{eq:triDP_scaling_2d_fl_asym}) 
by the reduced model parameter $\delta q$ if 
\begin{equation}
\delta p \ll \delta q \, .
\label{eq:scal_cond_02}
\end{equation} 
This condition Eq.\,(\ref{eq:scal_cond_02}) is fulfilled
if the simulations are performed in a way
that the distance to the phase boundary
is smaller than the distance to the tricritical point.
Thus, the order parameter and its 
fluctuations obey for $\delta q>0$ the scaling forms
\begin{eqnarray}
\label{eq:triDP_scaling_2d_op_pq}
\rhoa & \sim & \lambda^{-{\beta_{\scriptscriptstyle \mathrm{t}}}} \;
{\tilde r}_{\scriptscriptstyle {q}}
(\lambda \delta p,\,  \delta q \lambda^\phi,\,  h=0) \, , \\[2mm]
\label{eq:triDP_scaling_2d_fl_pq}
\Delta\rhoa & \sim & \lambda^{{\gamma^{\prime}_{\scriptscriptstyle \mathrm{t}}}} \;
{\tilde d}_{\scriptscriptstyle {q}}
(\lambda \delta p,\,  \delta q \lambda^\phi,\,  h=0)
\end{eqnarray}
if the critical point~$p_{\scriptscriptstyle \mathrm{c}}(q)$
is approached along $q=\mathrm{const}$ paths (as indicated by the index~$q$).
The above scaling forms are not valid
if $\nabla g \perp {\underline t}$ or if 
the derivative $\partial \tau/ \partial p$ vanishes
at the tricritical point.
In the latter case higher-orders 
($\mathcal{O}(\delta p^2)$) are required to describe the
scaling behavior.

In the next section we describe the analysis of the 
simulation data.
In particular, we have measured the averaged density of active sites
$\rho_{\scriptscriptstyle \mathrm a}(p,q,h)=
\langle L^{-\scriptscriptstyle D} 
N_{\scriptscriptstyle \mathrm a} \rangle$, i.e., the
order parameter as a function of~$p$ and~$q$
and of the conjugated field~$h$.
Numerically obtained order parameter 
curves for $q=0$ and for different values of the conjugated
field are plotted in Fig.\,\ref{fig:cp_rho_a_01}.
The conjugated field is implemented via a generation of 
particles ($0 \longrightarrow A$ with probability~$h$).
It results in a rounding of the zero field
curves and the order parameter behaves smoothly as a function 
of the control parameter for finite field values.
For $h\to 0$ we recover the non-analytical order parameter behavior.
Additionally to the order parameter, we investigate the
order parameter 
fluctuations $\Delta\rho_{\scriptscriptstyle \mathrm a}(p,g,h)
=L^{\scriptscriptstyle D}(
\langle \rho_{\scriptscriptstyle \mathrm a}^{\scriptscriptstyle 2} \rangle -
\langle \rho_{\scriptscriptstyle \mathrm a} \rangle^{\scriptscriptstyle 2})$
and the susceptibility~$\chi(p,q,h)=\partial \rhoa/\partial h$.
The susceptibility is obtained by performing the numerical 
derivative of the order parameter 
$\rho_{\scriptscriptstyle \mathrm a}$ with 
respect to the conjugated field~$h$.

\section{\textsf{SECOND-ORDER PHASE TRANSITION: \\ DIRECTED PERCOLATION BEHAVIOR}}

\begin{figure}[b]
  \centering
  \includegraphics[width=7.0cm,angle=0]{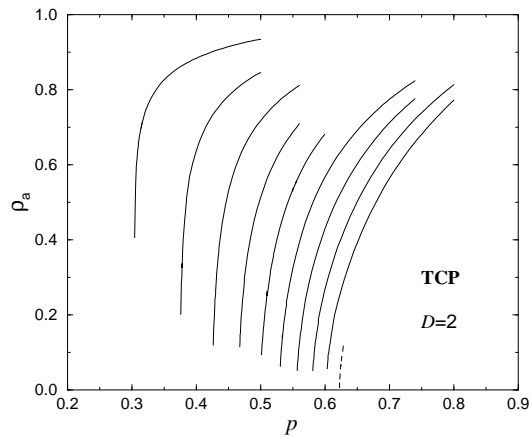}
  \caption{
    The order parameter~$\rhoa$ of the tricritical contact
    process (TCP) as a function of the control parameter~$p$
    for various values of~$q\in \{0,0.1,0.2,0.3,\ldots,0.9\}$ 
    (from right to left).
    The dashed line corresponds to the pure contact process ($q=0$).
    The data are obtained from simulations on simple cubic lattices
    of linear size~$L=64,128,\ldots,512$. 
    }
  \label{fig:tcp_rho_a_02} 
\end{figure}

\begin{figure}[t]
  \centering
  \includegraphics[width=7.0cm,angle=0]{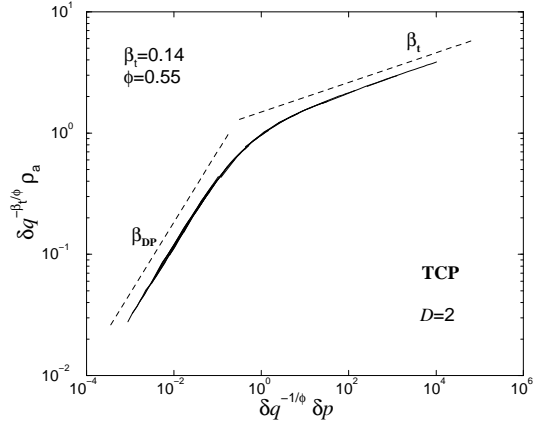}
  \caption{
    The crossover scaling function of the order parameter
    at zero field.
    The dashed lines correspond to the asymptotic
    behaviors, i.e., to the ordinary DP and to the 
    tricritical DP behavior 
    [see Eq.\,(\protect\ref{eq:triDP_scaling_2d_op_asym})].
    }
  \label{fig:cpt_crossover_01} 
\end{figure}

According to the above analysis we have simulated the 
tricritical contact process (TCP) and have measured~$\rhoa$
as a function of~$p$, keeping~$q$ fixed.
Order parameter curves for various $q$-value 
are shown in Fig.\,\ref{fig:tcp_rho_a_02}.
According to Eq.\,(\ref{eq:triDP_scaling_2d_op_pq})
these different curves
collapse onto a single curve if the rescaled
order parameter 
$\rhoa \, \delta q^{-\beta_{\scriptscriptstyle \mathrm{t}}/\phi}$
is plotted as a function of the rescaled
control parameter $\delta p \, \delta q^{-1/\phi}$.
Therefore, we vary the parameters
$\beta_{\scriptscriptstyle \mathrm{t}}$, $\phi$ 
as well as $q_{\scriptscriptstyle \mathrm{t}}$
until a data collapse is obtained.
Convincing results are obtained for 
$\beta_{\scriptscriptstyle \mathrm{t}}=0.14 \pm 0.02$, 
$\phi=0.55\pm0.03$ and 
$q_{\scriptscriptstyle \mathrm{t}}=0.9055\pm 0.0020$.
A corresponding scaling plot is shown 
in Fig.\,\ref{fig:cpt_crossover_01}.
The data are obtained from simulations for 16~different 
values of~$q$ ranging from $q=0.55$ up to $q=0.904$.
Typical distances to the critical line are of the order
of $\mathcal{O}(\delta p)=10^{-4}$.
On the other hand the minimal distance~$\delta q$ to the tricritical
point along the $q$-axis is larger than~$0.0015$.
In that way, the condition Eq.\,(\ref{eq:scal_cond_02}) is 
fulfilled, justifying the use of the scaling forms.

\begin{figure}[t]
  \centering
  \includegraphics[width=7.0cm,angle=0]{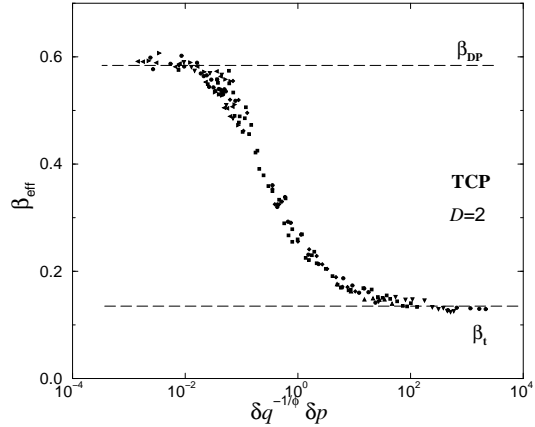}
  \caption{
    The effective 
    exponent $\beta_{\scriptscriptstyle \mathrm{eff}}$
    of the order parameter.
    Both asymptotic scaling
    regimes ($\beta_{\scriptscriptstyle \mathrm{t}}$
    and $\beta_{\scriptscriptstyle \mathrm{DP}}$) 
    as well as the crossover regime are clearly recovered.
    }
  \label{fig:beta_eff_02} 
\end{figure}
\begin{figure}[t]
  \centering
  \includegraphics[width=7.0cm,angle=0]{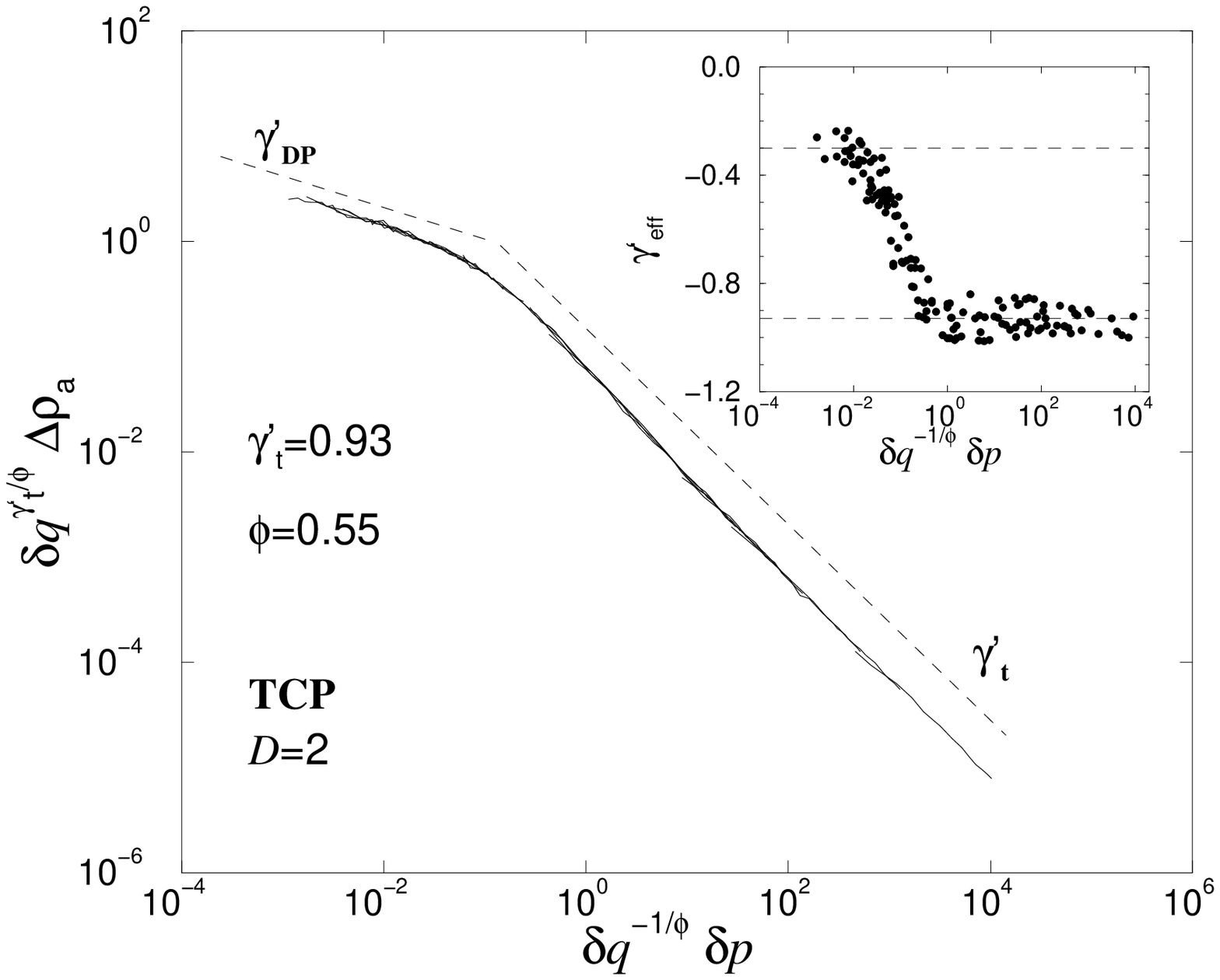}
  \caption{
    The crossover scaling function of the order 
    parameter fluctuations at zero field.
    The dashed lines correspond to the asymptotic
    behaviors, i.e., to the ordinary DP and to the 
    tricritical DP behavior 
    [see Eq.\,(\protect\ref{eq:triDP_scaling_2d_fl_asym})].
    The inset displays the corresponding effective 
    exponent $\gamma^{\prime}_{\scriptscriptstyle \mathrm{eff}}$.
    }
  \label{fig:cpt_crossover_02} 
\end{figure}

Since the entire crossover region covered several 
decades it could be difficult to observe small but systematic
differences between the scaling functions of both models.
It its therefore instructive to examine the crossover
via the so-called effective exponent
\begin{equation}
\beta_{\scriptscriptstyle \mathrm {eff}} \; = \;
\frac{\partial}{\partial \ln{x}}
\, \ln{{\tilde{r}_{q}}(x,1,0)} .
\label{eq:def_eff_exp_beta}
\end{equation}
The corresponding data are shown in Fig.\,\ref{fig:beta_eff_02}.
The excellent data collapse 
of~$\beta_{\scriptscriptstyle \mathrm {eff}}$ 
over more than 6 decades reflects the accuracy 
of the determination of the tricritical point.

A similar crossover scaling analysis can be performed for the 
order parameter fluctuations.
In that case, we use the above determined values of 
$\beta_{\scriptscriptstyle \mathrm{t}}$,
$q_{\scriptscriptstyle \mathrm{t}}$ and 
vary the value of the fluctuation 
exponent~$\gamma^{\prime}_{\scriptscriptstyle \mathrm{t}}$
until a data collapse of the different $q$-curves occurs.
Figure\,\ref{fig:cpt_crossover_02} shows the corresponding
data collapse as well as the effective exponent
\begin{equation}
\gamma^{\prime}_{\scriptscriptstyle \mathrm {eff}} \; = \;
\frac{\partial}{\partial \ln{x}}
\, \ln{{\tilde{d}_{q}}(x,1,0)} .
\label{eq:def_eff_exp_gammaprime}
\end{equation}
Although the data of the effective exponent 
are suffering from statistical fluctuations
both asymptotic scaling
regimes as well as the crossover regime can be identified.
Worth mentioning, the crossover scaling function exhibits a
so-called non-monotonic crossover~\cite{LUIJTEN_1}.
Unfortunately, the statistical
scattering of the effective exponent data masks the
non-monotonic behavior 
from~$\gamma^{\prime}_{\scriptscriptstyle \mathrm t}=0.93\pm 0.06$
to~$\gamma^{\prime}_{\scriptscriptstyle \mathrm{DP}}$.

In summary, the crossover scaling analysis allows the determination
of the tricritical value~$q_{\scriptscriptstyle \mathrm{t}}$
with high accuracy.
Furthermore, the effective 
exponents~$\beta_{\scriptscriptstyle \mathrm {eff}} $
and~$\gamma^{\prime}_{\scriptscriptstyle \mathrm {eff}}$
reflect the full crossover from tricritical DP to 
ordinary DP which spans more than 6 decades.

\section{\textsf{SECOND-ORDER PHASE TRANSITION: \\ TRICRITICAL BEHAVIOR}}

\begin{figure}[t]
  \centering
  \includegraphics[width=10.0cm,angle=0]{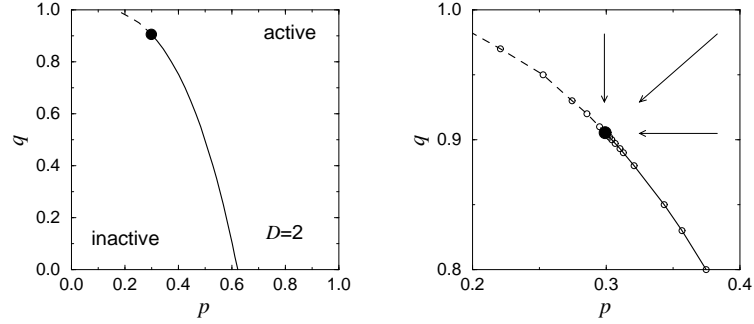}
  \caption{
    The phase diagram of the two-dimensional modified contact process.
    The solid line marks continuous phase transitions which
    belong to the universality class of directed percolation.
    The dashed line corresponds to first-order phase transitions
    and the bold circle indicates the tricritical point.
    The small circles in the right figure show where
    the transitions points are determined numerically.
    Furthermore, the tricritical point is approached in the
    simulations along three different ways illustrated by the 
    three arrows.
    }
  \label{fig:phasedia_2d} 
\end{figure}

In the previous section we have determined the tricritical value
$q_{\scriptscriptstyle \mathrm{t}}$
as well as the critical line $q(p_{\scriptscriptstyle \mathrm{c}})$
of the directed percolation like phase transitions.
This yield a first glance of the phase diagram which is
presented in Fig.\,\ref{fig:phasedia_2d}.
We now investigate the tricritical scaling behavior
in detail.
Therefore, we consider first the order parameter at zero field
in the vicinity of the tricritical point.
Second a conjugated field is applied which allows the
determination of various universal quantities such as 
tricritical exponents, scaling functions as well as amplitude combinations.

\begin{figure}[t]
  \centering
  \includegraphics[width=7.0cm,angle=0]{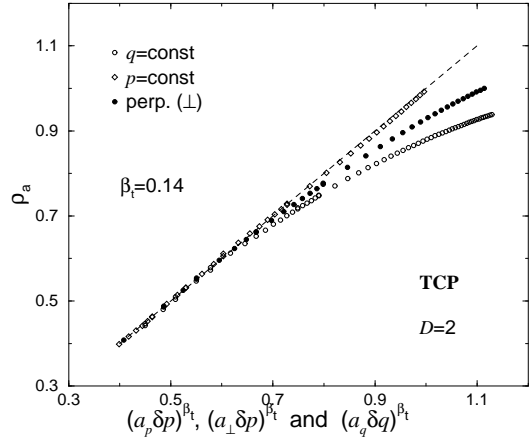}
  \caption{
    The order parameter behavior of the tricritical contact
    process (TCP). 
    The tricritical point is approached along
    three different ways indicated by the three different symbols. 
    All three curves tend
    asymptotically to the function $f(x)=x$ 
    (the dashed line corresponds to the pure
    tricritical power-law behavior) if  
    $\rhoa$ is plotted as a function of
    $(a_{\scriptscriptstyle p} \delta p)^{\beta_{\scriptscriptstyle \mathrm t}}$,
    $(a_{\scriptscriptstyle q} \delta q)^{\beta_{\scriptscriptstyle \mathrm t}}$,
    and 
    $(a_{\perp} \delta p)^{\beta_{\scriptscriptstyle \mathrm t}}$,
    respectively.
    But as can be seen the corrections to scaling depend
    on the scaling direction.
    }
  \label{fig:cpt_rho_a_02} 
\end{figure}

In the following analysis, the tricritical point is approached
in three different ways and the order parameter is determined at
zero field (see Fig.\,\ref{fig:phasedia_2d}).
First, $\rhoa$ is examined within the active phase
as a function of~$p$ at the tricritical value $q=0.9055$.
This yields the estimate 
\begin{equation}
p_{\scriptscriptstyle \mathrm{t}}=   
p_{\scriptscriptstyle \mathrm{c}}
(q_{\scriptscriptstyle \mathrm{t}}=0.9055)=0.29931\pm 0.00003\, .
\label{eq:estimate_pt}
\end{equation}
Second, the order parameter is measured for 
$p=0.29931$ as a function of~$q$.
From this analysis we obtain 
\begin{equation}
q_{\scriptscriptstyle \mathrm{c}}
(p_{\scriptscriptstyle \mathrm{t}}=0.29931)=0.90552\pm 0.00005 \, 
\label{eq:estimate_qt}
\end{equation}
which agrees with the above determined value 
$q_{\scriptscriptstyle \mathrm{t}}=0.9055\pm 0.0020$.
To check these estimates 
the critical point is approached along a third path which is
(more or less) perpendicular to the phase boundary.
As well known the leading order of the scaling behavior,
i.e.,~the critical exponents,
do not depend on the way the critical point is approached.
But the prefactors of the corresponding power-laws
and the corrections to the leading scaling order are affected
by the different directions.
Thus the scaling behavior of the order
parameter obeys asymptotically 
\begin{eqnarray}
\label{eq:op_tricrit_zf_qconst}
\rhoa & \sim &
(a_{\scriptscriptstyle p} \delta p)^{\beta_{\scriptscriptstyle \mathrm{t}}}
\quad \mathrm{for} \quad q=q_{\scriptscriptstyle \mathrm{t}}\\[2mm]
\label{eq:op_tricrit_zf_pconst}
\rhoa & \sim &
(a_{\scriptscriptstyle q} \delta q)^{\beta_{\scriptscriptstyle \mathrm{t}}}
\quad \mathrm{for} \quad p=p_{\scriptscriptstyle \mathrm{t}}\\[2mm]
\label{eq:op_tricrit_zf_perp}
\rhoa & \sim &
(a_{\scriptscriptstyle \perp} \delta p)^{\beta_{\scriptscriptstyle \mathrm{t}}}
\quad \mathrm{along~the~perpendicular~path.} 
\end{eqnarray}
Plotting the corresponding data accordingly 
(e.g. $\rhoa$ as a function of 
$(a_{\scriptscriptstyle p} \delta p)^{\beta_{\scriptscriptstyle \mathrm{t}}}$)
the three different curves collapse asymptotically 
to the leading tricritical behavior if the so-called non-universal
metric factors~$a_{\scriptscriptstyle p}$,
$a_{\scriptscriptstyle q}$,
$a_{\scriptscriptstyle \perp}$ are chosen appropriately.
The leading tricritical behavior corresponds in that analysis 
to a straight line with slope one.
As can be clearly seen in Fig.\,\ref{fig:cpt_rho_a_02}
all three curves approach the tricritical
scaling behavior asymptotically, confirming the accuracy of the
determination of the tricritical point and of the order parameter
exponent $\beta_{\scriptscriptstyle \mathrm{t}}$.
Furthermore, the corrections to scaling, i.e., the 
deviations to the asymptotical power-law depend strongly
on the way the tricritical point is approached.
Surprisingly, the smallest corrections occurs along the 
path $p=\mathrm{const}$ and not as usually expected along
the way which is perpendicular to the phase boundary.

So far we have considered the order parameter at zero field
in the vicinity of the tricritical point.
Applying an external field which is conjugated to the
order parameter, it is possible to investigate the 
scaling behavior of the tricritical equation of state.
Again, the conjugated field is implemented via a generation of 
particles ($0 \longrightarrow A$ with probability~$h$).
Sufficiently close to the tricritical point ($g=0$)
the order parameter obeys the scaling form
\begin{eqnarray}
\rhoa(\tau_{\scriptscriptstyle p,q},g_{\scriptscriptstyle p,q},h) 
& \sim & \left . \lambda^{-{\beta_{\scriptscriptstyle \mathrm{t}}}} \;
{\tilde r}
(\lambda \tau_{\scriptscriptstyle p,q},
\,  g \lambda^\phi,  \, 
h \lambda^{\sigma_{\scriptscriptstyle \mathrm{t}}^{\vphantom X}} \, 
)^{\vphantom X} \right |_{g=0} \nonumber \\[2mm]
& = & \lambda^{-{\beta_{\scriptscriptstyle \mathrm{t}}}} \;
{\tilde r}
(\lambda \tau_{\scriptscriptstyle p,q},0,  
h \lambda^{\sigma_{\scriptscriptstyle \mathrm{t}}} \, 
) \, .
\label{eq:triDP_scaling_2d_eqos_01}
\end{eqnarray}
Here, $\tau_{\scriptscriptstyle p,q}$ describes the distance to the
tricritical point within the $p-q-$plane.
All non-universal system dependent
features (such as the lattice structure or the way the tricritical 
point is approached, etc.) can be absorbed in two metric factors 
$a_{\scriptscriptstyle \mathrm{path}}$ and $a_{\scriptscriptstyle h}$. 
Once the non-universal metric factors are chosen in a specific
way, the scaling function ${\tilde R}$ (in contrast to
${\tilde r}$) is the same 
for all systems belonging to the universality class of directed
percolation.
Thus the universal scaling behavior is described by the ansatz
\begin{equation}
\rhoa(\tau_{\scriptscriptstyle p,q},0,h) 
\sim  \lambda^{-{\beta_{\scriptscriptstyle \mathrm{t}}}} \;
{\tilde R}
(\lambda a_{\scriptscriptstyle \mathrm{path}} 
\tau_{\scriptscriptstyle \mathrm{path}},0,  
a_{\scriptscriptstyle h}h \lambda^{\sigma_{\scriptscriptstyle \mathrm{t}}} 
) 
\label{eq:triDP_scaling_2d_eqos_02}
\end{equation} 
yielding for 
$a_{\scriptscriptstyle h}h \lambda^{\sigma_{\scriptscriptstyle \mathrm{t}}} =1$
\begin{equation}
\rhoa(\tau_{\scriptscriptstyle p,q},0,h) 
\sim   (a_{\scriptscriptstyle h}h )^{-{\beta_{\scriptscriptstyle \mathrm{t}}}
/\sigma_{\scriptscriptstyle \mathrm{t}}} \;
{\tilde R}
( a_{\scriptscriptstyle \mathrm{path}} 
\tau_{\scriptscriptstyle \mathrm{path}}
(a_{\scriptscriptstyle h} h)^{1/\sigma_{\scriptscriptstyle \mathrm{t}}},0,1) \, . 
\label{eq:triDP_scaling_2d_eqos_03}
\end{equation} 
Throughout this work we use the conditions 
${\tilde R}(1,0,0)={\tilde R}(0,0,1)=1$ to specify the 
metric factors which can be obtained from the 
amplitudes of the corresponding power-laws
[see e.g.~Eqs.\,(\ref{eq:op_tricrit_zf_qconst})-(\ref{eq:op_tricrit_zf_perp})].

\begin{figure}[t]
  \centering
  \includegraphics[width=7.0cm,angle=0]{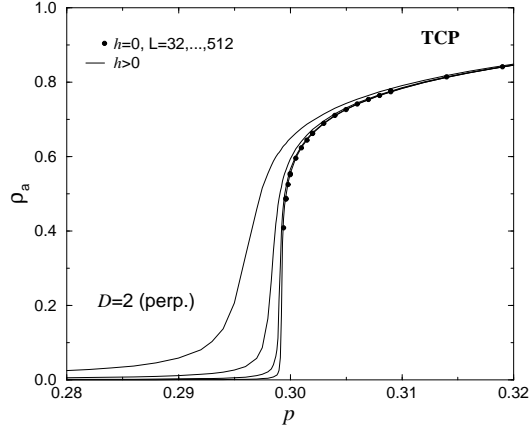}
  \caption{
    The order parameter~$\rhoa$ of the two-dimensional 
    tricritical contact process (TCP) close to the tricritical point.
    The tricritical point is crossed perpendicular to the 
    phase boundary (see Fig.\protect\ref{fig:phasedia_mf}).
    The order parameter is plotted as a function of the control parameter~$p$.
    The zero field data (circles) are obtained from simulations on a simple 
    cubic lattices of linear size~$L=64,128,256,512$. 
    The solid curves correspond to the non-zero field behavior
    (from $h=2\,10^{-3}$ to $h=3\,10^{-5}$).
    }
  \label{fig:cpt_rho_a_senk} 
\end{figure}

\begin{figure}[t]
  \centering
  \includegraphics[width=7.0cm,angle=0]{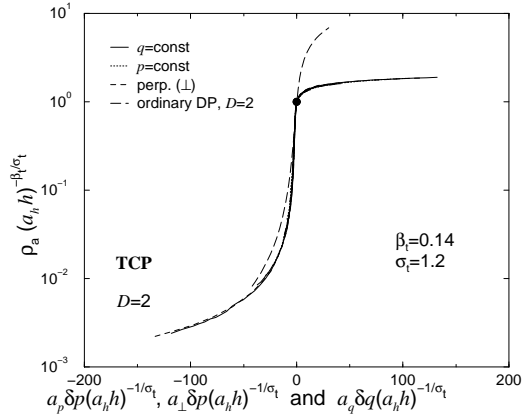}
  \caption{
    The universal scaling function 
    ${\tilde R}(x,0,1)$ of the 
    universality class of TDP.
    For all three scaling directions (only the data
    within the scaling regime are plotted) 
    at least four different curves are plotted
    corresponding to four different field values. 
    The circle marks the condition 
    ${\tilde R}(0,0,1)=1$.
    The dashed line corresponds to the universal scaling
    function of ordinary 
    DP (taken from~\protect\cite{LUEB_32}).
    }
  \label{fig:eqos_cpt_01} 
\end{figure}

In order to determine the universal scaling form of the equation
of state we have measured the field dependence of the order parameter.
Again the tricritical point is crossed along three different paths
in the $p-q-$plane.
Along each path, $\rhoa$ is determined as a function of the 
distance to the tricritical point for at least four
different field values.
The results are plotted for the perpendicular path in 
Fig.\,\ref{fig:cpt_rho_a_senk}.
According to the scaling form Eq.\,(\ref{eq:triDP_scaling_2d_eqos_03})
the rescaled order parameter is plotted as a function of the 
rescaled distances to the tricritical point.
In our analysis we have varied the field 
exponent~$\sigma_{\scriptscriptstyle \mathrm{t}}$ until a 
data collapse is obtained.
Convincing results are obtained for 
$\sigma_{\scriptscriptstyle \mathrm{t}}=1.20\pm 0.05$
and are shown in Fig.\,\ref{fig:eqos_cpt_01}.
Since data of different scaling directions are considered
the presented data collapse is
an impressive demonstration of the universality 
class of TDP.
Furthermore, the tricritical universal scaling
function differs significantly from the corresponding
scaling function of ordinary DP,
reflecting the different universality classes.

Additionally to the order parameter behavior we have
investigated the order parameter fluctuations~$\Delta\rhoa$ as well as
the order parameter susceptibility~$\chi$.
The scaling behavior of both quantities is described by~\cite{LUEB_32}
\begin{eqnarray}
\label{eq:triDP_scaling_2d_fluc_01}
a_{\Delta} \Delta \rhoa(\tau_{\scriptscriptstyle p,q},0,h) 
& \sim & \lambda^{{\gamma^{\prime}_{\scriptscriptstyle \mathrm{t}}}} \,
{\tilde D}
(\lambda a_{\scriptscriptstyle \mathrm{path}} 
\tau_{\scriptscriptstyle \mathrm{path}},0,  
a_{\scriptscriptstyle h}h \lambda^{\sigma_{\scriptscriptstyle \mathrm{t}}}), \\[2mm]
\label{eq:triDP_scaling_2d_susc_01}
a_{\chi} \chi(\tau_{\scriptscriptstyle p,q},0,h) 
& \sim & \lambda^{{\gamma_{\scriptscriptstyle \mathrm{t}}}} \;
{\tilde {\mathrm{X}}}
(\lambda a_{\scriptscriptstyle \mathrm{path}} 
\tau_{\scriptscriptstyle \mathrm{path}},0,  
a_{\scriptscriptstyle h}h \lambda^{\sigma_{\scriptscriptstyle \mathrm{t}}} 
) . 
\end{eqnarray} 
Setting ${\tilde D}(0,0,1)=1$ the non-universal metric 
factor~$a_{\scriptscriptstyle \Delta}$ is specified.
Taking into account that the susceptibility is defined as the derivative
of the order parameter with respect to the conjugated field
we find $a_{\scriptscriptstyle \chi}  =  a_{\scriptscriptstyle h}^{-1}$,
\begin{eqnarray}
\label{eq:sus_scal_func_01}
{\tilde {\mathrm X}}(x,0,y) & = & \partial_y \, {\tilde R}(x,0,y) ,\\[2mm]
\label{eq:sus_scal_func_02}
{\tilde {\mathrm{X}}}(0,0,1) & = &  
\frac{\,\beta_{\scriptscriptstyle \mathrm{t}}\,}
{\sigma_{\scriptscriptstyle \mathrm{t}}}  ,\\[2mm]
\label{eq:widom_law}
\gamma_{\scriptscriptstyle \mathrm{t}} & = &  
\sigma_{\scriptscriptstyle \mathrm{t}}   - 
 \beta_{\scriptscriptstyle \mathrm{t}} \, .
\end{eqnarray}
The latter equation corresponds to the Widom scaling law
in equilibrium.
Furthermore, Eq.\,(\ref{eq:sus_scal_func_02}) offers a useful
consistency check of the numerical analysis.

\begin{figure}[t]
  \centering
  \includegraphics[width=7.0cm,angle=0]{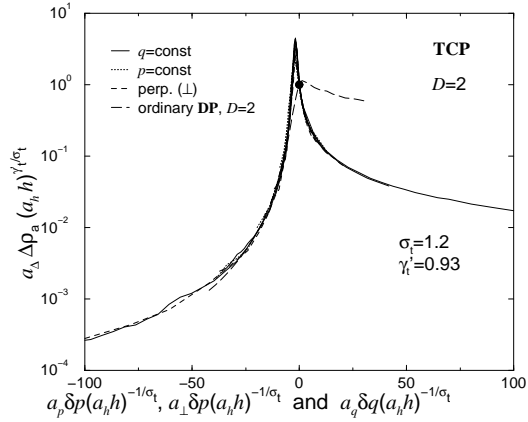}
  \caption{
    The universal scaling function 
    ${\tilde D}(x,0,1)$ of the
    universality class of tricritical directed percolation.
    For all three scaling directions (only the data
    within the scaling regime are plotted) the scaling
    plot contains at least four different curves
    corresponding to four different field values. 
    The circle marks the condition 
    ${\tilde D}(0,0,1)=1$
    }
  \label{fig:fluc_cpt_01} 
\end{figure}

\begin{figure}[t]
  \centering
  \includegraphics[width=7.0cm,angle=0]{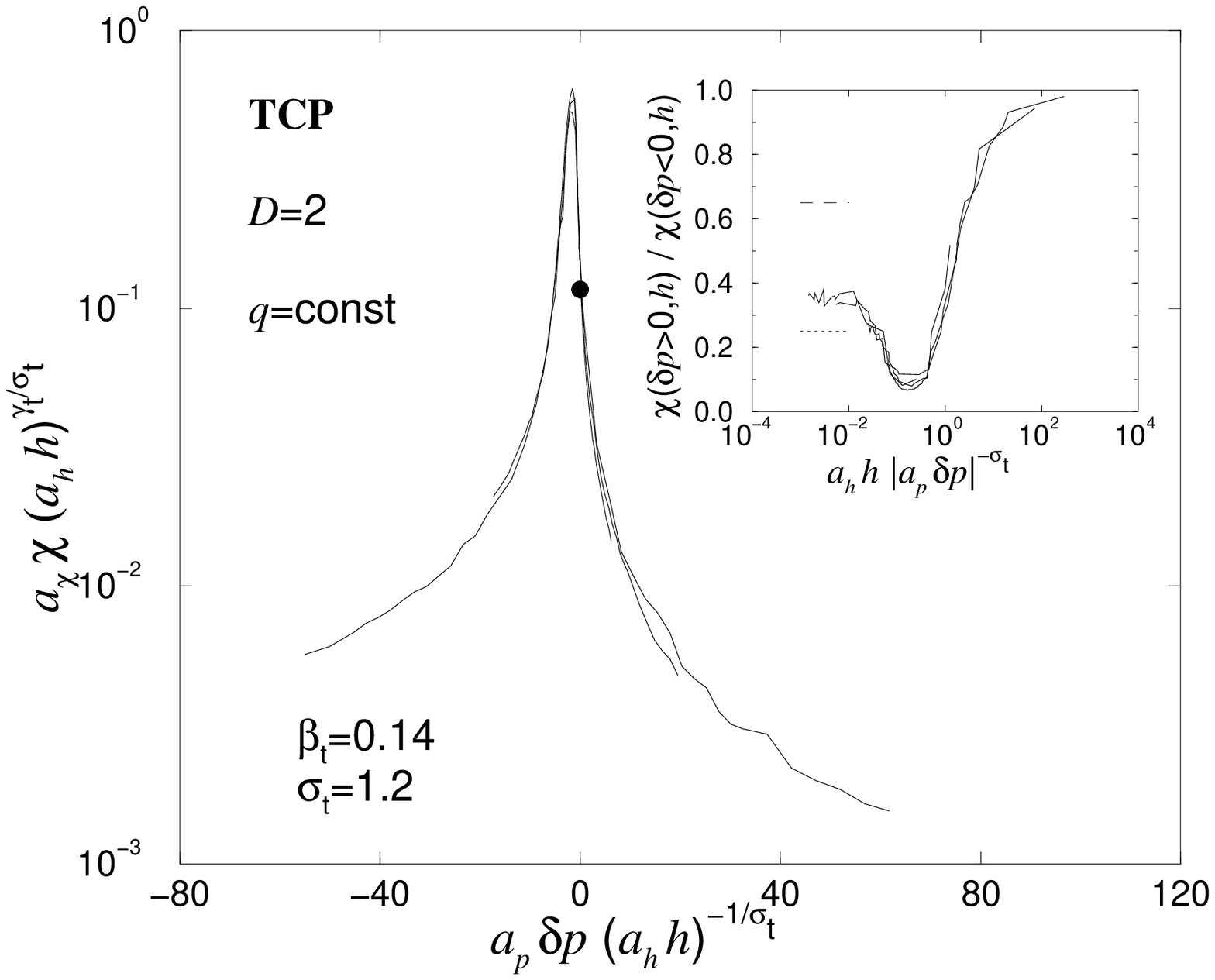}
  \caption{
    The universal scaling function 
    ${\tilde \mathrm{X}}(x,0,1)$ of the
    universality class of tricritical directed percolation.
    The circle marks the condition 
    ${\tilde \mathrm{X}}(0,1,0)=
    \beta_{\scriptscriptstyle \mathrm{t}}/
    \sigma_{\scriptscriptstyle \mathrm{t}}$ and reflects
    the accuracy of the performed analysis.
    The inset displays the universal ratio
    ${\tilde \mathrm{X}}(1,0,x)/{\tilde \mathrm{X}}(-1,0,x)$.
    The extrapolation ($x\to 0$)
    yields the value of the universal amplitude ratio 
    ${\tilde \mathrm{X}}(1,0,0)
    /{\tilde \mathrm{X}}(-1,0,0)=0.35\pm0.04$.
    The obtained value differs significantly from the
    corresponding values (dashed lines) of ordinary directed
    percolation $0.25$ for $D=2$ and $0.65$ for $D=3$~\cite{LUEB_28}, 
    respectively.
    }
  \label{fig:susc_cpt_01} 
\end{figure}

The universal scaling functions of the order parameter
fluctuations and the order parameter susceptibility are shown
in Fig.\,\ref{fig:fluc_cpt_01} and Fig.\,\ref{fig:susc_cpt_01}, 
respectively.
The susceptibility is obtained by performing the numerical
derivative of the order parameter with respect to the 
conjugated field.
Both scaling functions exhibit a maximum signaling the 
divergence of $\Delta\rhoa$ and $\chi$ at the tricritical
point.
The susceptibility data fulfill Eq.\,(\ref{eq:sus_scal_func_02}),
reflecting the accuracy of the performed analysis.
In both cases, the obtained universal scaling functions differ
significantly from the corresponding scaling functions
of ordinary DP.

Additionally to the critical exponents and universal scaling
functions it is useful to investigate universal
amplitude combinations.
An often considered amplitude combination is related to
the susceptibility behavior below and above the 
transition.
For e.g. $q=\mathrm{const}$ the susceptibility diverges as
\begin{eqnarray}
\chi(\delta p >0,g=0,h=0) & \sim &
a_{\scriptscriptstyle \chi,+} \; 
\delta p^{-\gamma_{\scriptscriptstyle \mathrm{t}}} \, , \\[2mm]
\chi(\delta p <0,g=0,h=0) & \sim &
a_{\scriptscriptstyle \chi,-} \; 
(-\delta p)^{-\gamma_{\scriptscriptstyle \mathrm{t}}} \, ,
\end{eqnarray}
if the critical point is approached from above
and below, respectively.
Using Eq.\,(\ref{eq:triDP_scaling_2d_susc_01})
the susceptibility ratio
\begin{eqnarray}
\label{eq:uni_ampl_comb_sus_X_x1}
\frac{\,\chi(\delta p >0,0,h)\,}{\chi(\delta p <0,0,h)}
& = &
\left .
\frac{\,{\tilde \mathrm{X}}
(\phantom{-}a_{\scriptscriptstyle p} \delta p \; \lambda,0, 
a_{\scriptscriptstyle h} h \, \lambda^{\sigma_{\scriptscriptstyle \mathrm{t}}})\,}
{\,{\tilde \mathrm{X}}
(-a_{\scriptscriptstyle p} \delta p \; \lambda,0,
a_{\scriptscriptstyle h} h \, \lambda^{\sigma_{\scriptscriptstyle \mathrm{t}}})\,}
\, \right |_{a_{\scriptscriptstyle p} |\delta p| \lambda=1} 
\nonumber \\[2mm]
& = & 
\frac{\,{\tilde \mathrm{X}}
(+1,0,x)\,}
{\,{\tilde \mathrm{X}}(-1,0,x)\,}
\end{eqnarray}
is clearly a universal quantity for all values of the
scaling argument 
$x=a_{\scriptscriptstyle h} h |a_{\scriptscriptstyle p} 
\delta p|^{-\sigma_{\scriptscriptstyle \mathrm{t}}}$.
In particular it equals the 
ratio ${a_{\scriptscriptstyle \chi,+}}/{a_{\scriptscriptstyle \chi,-}}$
for vanishing field
\begin{equation}
\frac{\, a_{\scriptscriptstyle \chi,+}\, }{a_{\scriptscriptstyle \chi,-}}
\; = \;
\frac{\,{\tilde \mathrm{X}}(+1,0,0)\,}{\,{\tilde \mathrm{X}}(-1,0,0)\,} \, .
\label{eq:uni_ampl_comb_sus}
\end{equation}
For example, the mean field value of the universal ratio
equals $1/2$.

The ratio 
${\tilde \mathrm{X}}(1,0,x)
/{\tilde \mathrm{X}}(-1,0,x)$ is shown in the inset
of Fig.\,\ref{fig:susc_cpt_01}.
The extrapolation to the tricritical point ($x\to 0$)
yields the value of the universal
amplitude ratio ${\tilde \mathrm{X}}(1,0,0)
/{\tilde \mathrm{X}}(-1,0,0)=0.35\pm0.05$.
This value differs significantly from the
corresponding values of ordinary DP for $D=2$ and $D=3$ 
(see Fig.\,\ref{fig:susc_cpt_01} as well as Table\,\ref{table:exponents}).
Worth mentioning, the universal susceptibility ratio
exhibits a non-monotonic behavior as a function of the 
scaling argument~$x$.
This non-monotonic behavior is a characteristic
feature of TDP
and does not occur in ordinary~DP 
(see e.g.\,Fig.\,30 of \protect\cite{LUEB_35}).

So far simulation data are taken into account where the correlation 
length is small compared to the system size~$L$.
Thus, the data presented above do not suffer from 
finite-size effects, such as rounding and shifting of the anomalies.
A typical feature of finite-size effects in equilibrium
is that a given system may pass within the simulations from one 
phase to the other.
This behavior is caused by critical fluctuations which increase
if one approaches the transition point.
In case of absorbing phase transitions the scenario is different.
Approaching the transition point, the (spatial) correlation 
length~$\xi_{\scriptscriptstyle \perp}$ 
increases.
As soon as~$\xi_{\scriptscriptstyle \perp}$ 
is of the order of~$L$ the system may pass to the absorbing
state where it is trapped forever.
As pointed out in~\cite{LUEB_23} an appropriate way to handle
that problem is to incorporate a conjugated field.
Due to the conjugated field the system can not be trapped 
within the absorbing state and steady state quantities are
available for all values of the control parameter.

As usual the system size~$L$ enters the scaling forms as an
additional scaling field, e.g.
\begin{equation}
\rhoa
\sim  \lambda^{-{\beta_{\scriptscriptstyle \mathrm{t}}}} \;
{\tilde R}
(\lambda a_{\scriptscriptstyle \mathrm{path}} 
\tau_{\scriptscriptstyle \mathrm{path}},0,  
a_{\scriptscriptstyle h}h \lambda^{\sigma_{\scriptscriptstyle \mathrm{t}}},
a_{\scriptscriptstyle L}L \lambda^{-\nu_{\scriptscriptstyle \perp,\mathrm{t}}}
) 
\label{eq:triDP_scaling_2d_eqos_fss}
\end{equation} 
with the tricritical 
exponent~$\nu_{\scriptscriptstyle \perp,\mathrm{t}}$ 
of the spatial correlation length.
Using the above scaling form it is possible to determine
the correlation length exponent from data of different system sizes.
As well known from equilibrium, ratios of order 
parameter moments~$\langle \rhoa^k \rangle$
are more suited to estimate the correlation
length exponent.
For example, the well-known Binder cumulant  
$Q=1-\langle \rhoa^4 \rangle/3 \langle \rhoa^2 \rangle$ 
was successfully investigated in numerous works
(see e.g.~\cite{BINDER_1}) dealing with equilibrium as well 
as non-equilibrium phase transitions.
Furthermore, the value of the Binder cumulant at the transition 
point is a universal quantity.

\begin{figure}[t]
  \centering
  \includegraphics[width=7.0cm,angle=0]{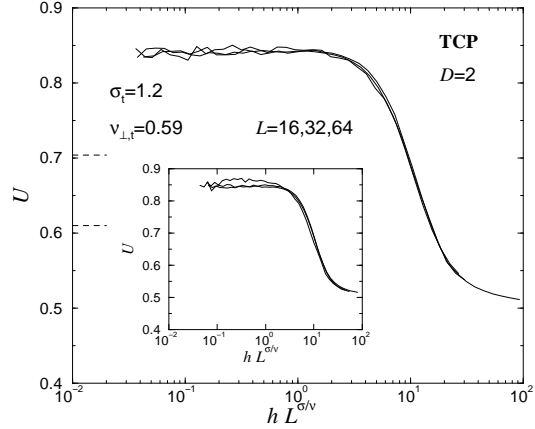}
  \caption{
    Scaling plot of the ratio $U(h,L)$ close to the tricritical point
    ($p=0.29931$,$q=0.90552$).
    For vanishing field the ratio tends to the 
    universal value $U=0.84\pm0.04$ which characterizes
    the universality class.
    The dashed lines correspond to the values of~$U(h\to 0)$
    for two-dimensional and three-dimensional directed
    percolation.
    The inset displays the ratio slightly away from the
    critical point ($p=0.29934$,$q=0.90557$).
    The systematic deviations from a data collapse 
    for small values of the scaling argument
    reflect how sensible~$U$ depends
    on the determination of the critical point.
    }
  \label{fig:U_cpt_01} 
\end{figure}

Unfortunately, the Binder cumulant diverges at the critical
point of absorbing phase transitions~\cite{LUEB_23,LUEB_33}.
This behavior is caused by the vanishing steady state
fluctuations in the absorbing phase and 
reflects the different nature of the zero-order
parameter phase in equilibrium and in absorbing phase
transitions.
A ratio that remains finite at criticality is given 
by~\cite{LUEB_33}
\begin{equation}
U \; = \; 
\frac{\langle\rhoa^{2}\rangle  \langle\rhoa^{3}\rangle \,  - \,
\langle\rhoa\rangle   \langle\rhoa^{2}\rangle^{2}}
{\langle\rhoa\rangle \langle\rhoa^{4}\rangle \, -\, 
\langle\rhoa\rangle \, \langle\rhoa^{2}\rangle^{2}} \, .
\label{eq:def_ratio_U}
\end{equation}
This ratio is as useful for absorbing phase transitions
as the Binder cumulant~$Q$ is for equilibrium, i.e.,
its value at criticality characterizes the universality class.
Here, we investigated the ratio~$U$ close to the
tricritical point ($\tau=0$ and $g=0$). 
Its scaling behavior obeys
\begin{eqnarray}
U & =  & 
\left . {\tilde U}(
\lambda 
\tau_{\scriptscriptstyle p,q}, 
g_{\scriptscriptstyle p,q} \lambda^{\phi},  
h \lambda^{\sigma_{\scriptscriptstyle \mathrm{t}}},
L 
\lambda^{-\nu_{\scriptscriptstyle \perp,\mathrm{t}}}
) \right |_{p_{\scriptscriptstyle \mathrm{t}},q_{\scriptscriptstyle \mathrm{t}}}
\nonumber \\[2mm] 
& = & \left . {\tilde U}(0,0,
h \lambda^{\sigma_{\scriptscriptstyle \mathrm{t}}},
L \lambda^{-\nu_{\scriptscriptstyle \perp,\mathrm{t}}}
) \right |_{L
\lambda^{-\nu_{\scriptscriptstyle \perp,\mathrm{t}}}=1 }
\nonumber \\[2mm] 
\label{eq:triDP_scal_U}
& = &  {\tilde U}(0,0,
h L^{\sigma_{\scriptscriptstyle\mathrm{t}}/
\nu_{\scriptscriptstyle \perp,\mathrm{t}}}
) \, .
\end{eqnarray} 
The ratio $U$ is shown in Fig.\,\ref{fig:U_cpt_01}
for $p=0.29931$ and $q=0.90552$.
Convincing data collapses are obtained for 
$\nu_{\scriptscriptstyle \perp,\mathrm{t}}=0.59\pm 0.08$.
As can be seen the ratio tends to a well defined value
for $h\to 0$ independent of the system size~$L$.
The obtained value $U=0.84\pm0.04$ differs significantly from the
corresponding values of two-dimensional 
and three-dimensional ordinary DP (see Table\,\ref{table:exponents}).
Note that the determination of $U(h\to 0)$ does not depend
on the critical exponents $\sigma_{\scriptscriptstyle \mathrm{t}}$
and $\nu_{\scriptscriptstyle \perp,\mathrm{t}}$.
But it is very sensitive to the determination of the
critical point.
Performing simulations slightly away from the critical
point the ratio $U(h\to 0)$ displays a clear system 
size dependence (see inset of Fig.\,\ref{fig:U_cpt_01}).



Thus we have determined the steady state scaling behavior 
of tricritical directed percolation.
The obtained values of the critical exponents as well as of the
universal amplitude ratios are listed in Table\,\ref{table:exponents}.
The accuracy of the estimated exponents can be checked
with the scaling law
\begin{equation}
\gamma_{\scriptscriptstyle {\mathrm{t}}}^{\prime} \; = \;
\nu_{\scriptscriptstyle \perp,{\mathrm{t}}} 
D \, - \, 2 \beta_{\scriptscriptstyle \perp,{\mathrm{t}}} \, .
\label{eq:scal_law_fluc}
\end{equation}
The determined two-dimensional  
values~$\gamma_{\scriptscriptstyle {\mathrm{t}}}^{\prime}=0.93\pm0.06$,
$\nu_{\scriptscriptstyle \perp,{\mathrm{t}}}=0.59\pm0.08$, and
$\beta_{\scriptscriptstyle {\mathrm{t}}}=0.14\pm0.02$
fulfill the above scaling law within the error bars.
Furthermore, the exponent of the spatial correlation 
function~$\eta_{\scriptscriptstyle \perp,{\mathrm{t}}}$
is related to the correlation length exponent and to the fluctuation
exponent via the Fisher scaling law
\begin{equation}
(2-\eta_{\scriptscriptstyle \perp,{\mathrm{t}}}) 
\nu_{\scriptscriptstyle \perp,{\mathrm{t}}} \; = \;
\gamma_{\scriptscriptstyle {\mathrm{t}}}^{\prime} \,  ,
\label{eq:scal_law_fisher}
\end{equation}
leading to 
$\eta_{\scriptscriptstyle \perp,{\mathrm{t}}}=0.42\pm0.24$.
Another quantity of interest is the fractal 
dimension~$D_{\scriptscriptstyle \mathrm{f}}$ of growing
clusters at criticality.
The fractal dimension is given by~\cite{GRASSBERGER_8,LUEB_35}
\begin{equation}
D_{\scriptscriptstyle \mathrm{f}} \; = \;
D \, - \, \frac{\beta_{\scriptscriptstyle \mathrm{t}}}
{\nu_{\scriptscriptstyle \perp,\mathrm{t}}}
\label{eq:scal_law_frac_dim}
\end{equation}
yielding $D_{\scriptscriptstyle \mathrm{f}}=1.76\pm 0.05$.
This value is larger than the corresponding values
of ordinary directed percolation 
($D_{\scriptscriptstyle \mathrm{f},D=2}\approx 1.20$
and 
$D_{\scriptscriptstyle \mathrm{f},D=3}\approx 1.60$).
A detailed investigation of the fractal behavior
of critical clusters is desirable.
For example, a determination of the lacunarity along
the phase boundary would present a deeper understanding
of the cluster propagation (see e.g.~\cite{HINRICHSEN_1,LEE_3}).

It is worth comparing our numerical results to those of
corresponding field theoretical analyses.
Within a two-loop approach the critical exponents
are given in linear order 
of $\epsilon=D_{\scriptscriptstyle \mathrm{c}}-D$ 
by~\cite{OHTSUKI_1,OHTSUKI_2,JANSSEN_4} 
\begin{eqnarray}
\label{eq:tdp_eps_beta}
\beta_{\scriptscriptstyle {\mathrm{t}}}
& = & \frac{1}{2} \, - \, \epsilon \, 0.4580\ldots \, , \\[2mm] 
\label{eq:tdp_eps_betap}
\beta^{\prime}_{\scriptscriptstyle {\mathrm{t}}} 
& = & 1 \, + \, {\mathcal{O}}(\epsilon^2) \, \\[2mm] 
\label{eq:tdp_eps_z}
z_{\scriptscriptstyle {\mathrm{t}}} 
& = & 2 \, + \, \epsilon\, 0.0086\ldots \, , \\[2mm] 
\label{eq:tdp_eps_nuperp}
\nu_{\scriptscriptstyle \perp,\mathrm{t}} 
& = & \frac{1}{2} \, + \, \epsilon \, 0.0075\ldots
\, , \\[2mm]
\label{eq:tdp_eps_gamma}
\gamma_{\scriptscriptstyle {\mathrm{t}}}
& = & 1 \, + {\mathcal{O}}(\epsilon^2) \, \\[2mm]
\label{eq:tdp_eps_gammap}
\gamma^{\prime}_{\scriptscriptstyle {\mathrm{t}}} 
& = & \frac{1}{2} \, + \, \epsilon\, 0.4386\ldots\, , \\[2mm]
\label{eq:tdp_eps_phi}
\phi
& = & \frac{1}{2} \, - \epsilon \, 0.0121\ldots\, .
\end{eqnarray}
Here, $\beta^{\prime}$ denotes the critical exponent of the 
survival probability.
According to the $\epsilon$-expansion one expects that the
two-dimensional ($\epsilon=1$) values of the 
exponents $\nu_{\scriptscriptstyle \perp,{\mathrm{t}}}$,
$\gamma_{\scriptscriptstyle {\mathrm{t}}}$, 
and $\phi$
differ only slightly from their mean field values.
Whereas strong deviations are predicted by the
$\epsilon$-expansion for
$\beta_{\scriptscriptstyle \mathrm{t}}$
and $\gamma_{\scriptscriptstyle \mathrm{t}}^{\prime}$.
This behavior is confirmed by our numerical
results.
But one has to mention that the field theorically estimated
exponents differ significantly from the numerical values.
For example, the $\epsilon$-expansion yields
for the order parameter exponent
$\beta_{\scriptscriptstyle \mathrm{t}}=0.042\ldots$.
This value differs by $70\%$ from the numerical value.
Furthermore, the RG results predict
$\phi< 1/2$ whereas the simulations
clearly show 
$\phi>1/2$.
Thus an $\epsilon$-expansion of higher-orders 
than $\mathcal{O}(\epsilon)$ is desirable to describe 
the scaling behavior by a field theoretical approach.

\begin{table}[t]
\centering
\caption{
The critical exponents and various universal amplitude combinations
of tricritical (TDP) 
and ordinary directed percolation (DP) for various dimensions~$D$.
The crossover exponent from tricritical to ordinary DP is
given by $\phi_{\scriptscriptstyle D=2}=0.55\pm 0.03$ 
and $\phi_{\scriptscriptstyle \mathrm{MF}}=1/2$.
}
\vspace{0.3cm}
\label{table:exponents}
\begin{tabular}{|c|l|c|c|c|c|}
\hline
$$       
& $\mathrm{TDP}_{\scriptscriptstyle D=2}$  
& $\mathrm{TDP}_{\scriptscriptstyle D>3}$  
& $\mathrm{DP}_{\scriptscriptstyle D=2}$ 
{\protect\cite{VOIGT_1,GRASSBERGER_3,LUEB_28,LUEB_33}}
& $D=3$ {\protect\cite{JENSEN_6,LUEB_28,LUEB_33}}
& $\mathrm{DP}_{\scriptscriptstyle D>4}$\\  
\hline
$\beta$    			&  $0.14\pm0.02$	&  $1/2$	
& $0.5834\pm0.0030$ 		&  $0.813\pm0.009$ 	&  $1$\\   
$\nu_{\perp}$     		&  $0.59\pm0.08$	&  $1/2$	
& $0.7333\pm0.0075$ 		&  $0.584\pm0.005$	&  $1/2$\\    
$\nu_{\protect\parallel}$  	&  $$			&  $1$	
& $1.2950\pm0.0060$ 		&  $1.110\pm0.010$ 	&  $1$\\    
$\sigma$     			&  $1.12\pm0.05$	&  $3/2$ 
& $2.1782\pm0.0171$ 		&  $2.049\pm0.026$ 	&  $2$\\    
$\protect\gamma^{\prime}$     	&  $0.93\pm0.06$	&  $1/2$
& $0.2998\pm0.0162$ 		&  $0.126\pm0.023$	&  $0$\\   
$\gamma$        		&  $1.00\pm0.06$	&  $1$ &
$1.5948\pm0.0184$ 		&  $1.237\pm 0.023$ 	&  $1$\\  
$\eta_{\perp}$     		&  $0.42\pm0.24$	&  $1$	
& $1.5912\pm0.0148$ 		&  $1.783\pm0.016$	&  $2$\\    
$D_{\mathrm{f}}$     		&  $1.76\pm0.05$	&  $2$	
& $1.2044\pm0.0091$ 		&  $1.608\pm0.019$	&  $2$\\    
\hline
$\beta^{\prime}$		& $$			&  $1$	
& $\beta=\beta^{\prime}$ 	& $\beta=\beta^{\prime}$ 	&  $1$\\   
$\delta$ 			& $$ 			&  $1$	    
& $0.4505\pm0.0010$   		& $0.732\pm 0.004$ 	&  $1$\\ 
$\alpha$ 			& $$ 			&  $1/2$	    
& $\alpha=\delta$   		& $\alpha=\delta$ 	&  $1$\\ 
$\theta$      			& $$   			&  $0$  
& $0.2295\pm0.0010$  		& $0.114\pm 0.004$ 	&  $0$\\
$z$ 	      			& $$	  		&  $2$
& $1.7660\pm0.0016$  		& $1.901\pm0.005$ 	&  $2$\\ 
\hline
$\frac{{\tilde \mathrm{X}}(+1,0)}{{\tilde \mathrm{X}}(-1,0)}$ 
&  $0.35\pm0.05$   &  $1/2$	    
& $0.25\pm0.01$   		& $0.65\pm0.03$ 	& $1$\\      
$U$ 
&  $0.84\pm0.04$  &  $$	    
& $0.704\pm0.013$   		& $0.61\pm0.02$ 	& $1/2$\\ 
\hline
\end{tabular}
\end{table}

\section{\textsf{FIRST-ORDER PHASE TRANSITION}}

\begin{figure}[t]
  \centering
  \includegraphics[width=7.0cm,angle=0]{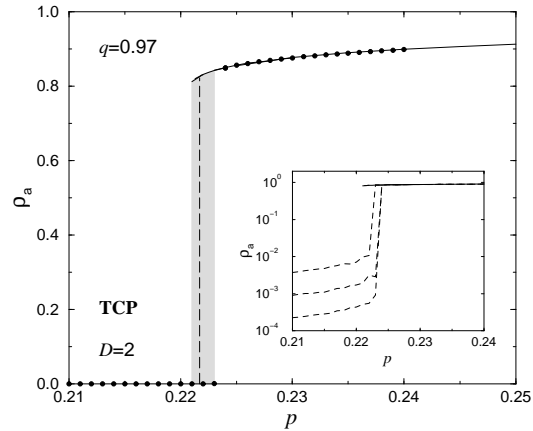}
  \caption{
    The order parameter behavior within the first-order
    regime of the tricritical contact process (TCP).
    The solid line is obtained from steady state  
    measurements at zero field for various system sizes.
    Within the simulations the control parameter~$p$
    is slowly decreased until the absorbing state is reached.
    The circles correspond to data which are
    obtained from finite-field simulations and slowly
    increasing parameter~$p$.
    Within the shadowed area the active and the absorbing
    phase coexist.
    The dashed line marks the transition 
    point~$p_{\scriptscriptstyle \mathrm{c}}$ obtained from a
    stability analysis of separated phases (see text).
    The inset shows the field dependence of the 
    order parameter (for growing~$p$). 
    The three dashed lines correspond to three
    different field values (from $h=4\,10^{-4}$ (top) to 
    $h=3\,10^{-5}$).
    As can be seen, a well defined upper limit of the
    supercooled low density phase can be obtained.
    }
  \label{fig:hysterese_01} 
\end{figure}

In this section we investigate the first-order regime of the 
modified contact process.
A general phenomenon associated with first-order transitions
is the presence of hysteresis by cycling across the transition.
In equilibrium, the hysteresis is related to the effects of
supercooling and superheating.
Analogous effects occur in case of the modified contact process.
First we investigate the effect of superheating, i.e.,
we consider the order parameter within the active phase ($\rhoa>0$)
while approaching the transition point (on heating in equilibrium).
Decreasing the parameter~$p$ for fixed 
$q>q_{\scriptscriptstyle \mathrm{t}}$ the order
parameter jumps at a certain point $p_{\scriptscriptstyle \mathrm{o}}$ 
from a finite value to zero.
This is shown in Fig.\,\ref{fig:hysterese_01} for $q=0.97$.
Close to $p_{\scriptscriptstyle \mathrm{o}}=0.221$
the order parameter behaves discontinuously.
This transition point $p_{\scriptscriptstyle \mathrm{o}}$
corresponds to the limit of superheating.
Note that in contrast to a continuous phase transition the 
transition point $p_{\scriptscriptstyle \mathrm{o}}$ exhibits no
systematic system size dependence.

The similar phenomenon of supercooling the zero-order parameter 
phase is usually not accessible for absorbing phase transitions.
Owing to the lack of fluctuations for $\rhoa=0$ 
the system can never escape the absorbing state
by a variation of~$p$ and~$q$, respectively.
To avoid that the system is trapped within the 
absorbing state an external field~$h$ is applied.
In that way it is possible to perform steady state measurements
of the order parameter on cooling, i.e., 
within in the low density phase ($\rhoa\ll 1$) for increasing~$p$.
The resulting curves are shown in the inset of 
Fig.\,\ref{fig:hysterese_01}.
At a certain value~$p_{\scriptscriptstyle \mathrm{u}}(h)$
the order parameter jumps from a low density value to
a high density value.
Note that the low density value of $\rhoa$ 
tends to zero for vanishing field.
Furthermore, a well defined transition value
$p_{\scriptscriptstyle \mathrm{u}}(h)$ exists for $h\to 0$.
The obtained value $p_{\scriptscriptstyle \mathrm{u}}\approx 0.223$
corresponds to the limit of supercooling.

\begin{figure}[t]
  \centering
  \includegraphics[width=9.0cm,angle=0]{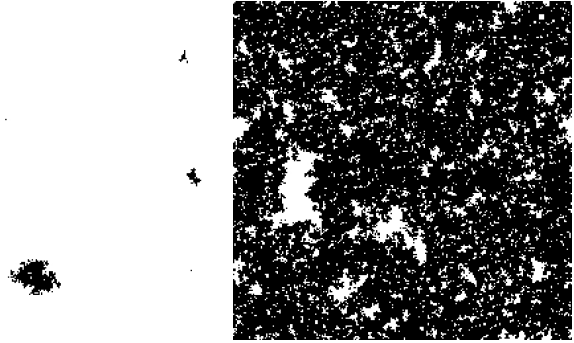}
  \caption{
    Snapshots of the tricritical contact process within the 
    regime of first-order phase transitions ($q=0.97$ and $L=256$).
    The left figure shows a typical low density configuration 
    ($p=0.223$, supercooled phase).
    The seeds are triggered by an external field $h=10^{-4}$.
    The right figure displays a high density configuration
    ($p=0.221$, superheated phase).
    Here, order parameter fluctuations lead to seeds of various 
    sizes.
    In both cases the seeds are subcritical, i.e., they 
    disappear after a certain lifetime.
    }
  \label{fig:snapshot_hyst} 
\end{figure}

In that way we have obtained from steady state measurements 
a small but finite hysteresis, i.e., the 
two phases coexist between  
$p_{\scriptscriptstyle \mathrm{o}}<p<p_{\scriptscriptstyle \mathrm{u}}$.
Within the active phase ($\rhoa\approx 0.8$) the system
is stable against small fluctuations until 
$p_{\scriptscriptstyle \mathrm{o}}$ is reached from above.
On the other hand the absorbing phase is stable 
against external fluctuations, triggered by the 
conjugated field, until $p_{\scriptscriptstyle \mathrm{u}}$
is approached from below.
Snapshots of the system within the supercooled and superheated state
are shown in Fig.\,\ref{fig:snapshot_hyst}.

Finally we address the question of the critical 
value~$p_{\scriptscriptstyle \mathrm{c}}(q)$
of the first-order phase transitions.
In equilibrium, the transition point is related to a 
thermodynamical potential such as the free energy.
At the critical temperature
the free energy of both phases are equal.
Unfortunately, this definition can not be applied
to the considered non-equilibrium phase transition.
An alternative way of defining the first-order transition point
is based on the behavior of moving interfaces
which separate both phases.
In case of phase equilibrium the interface velocity
is zero whereas it is non-zero if one phase is
favored by the dynamics.

\begin{figure}[t]
  \centering
  \includegraphics[width=6.0cm,angle=0]{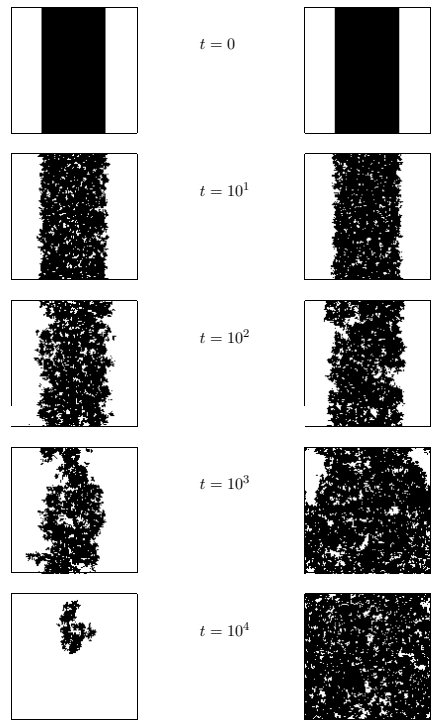}
  \caption{
    Snapshots of the tricritical contact process close
    to the first-order transition for $q=0.97$ and $L=128$
    (periodic boundary conditions are applied).
    Starting from a stripe of particles the dynamics
    is attracted 
    either by the empty lattice (left, $p=0.22$, the eventual
    occurring empty lattice is not shown) or 
    by a steady state of a homogeneous non-zero
    particle density (right, $p=0.24$).
    The time~$t$ is measured as the number of lattice updates.
    }
  \label{fig:wand_dyn} 
\end{figure}

According to that picture we have investigated the
phase propagation within the first-order regime.
Initially the system contains a stripe (width $L/2$)
of occupied particles. 
All lattice site outside the stripe remains empty.
Depending on~$p$ and~$q$,
the system reaches after a transient either the absorbing
phase or a steady state of a homogeneous non-zero
particle density.
Snapshots are shown in Fig.\,\ref{fig:wand_dyn}.
For $q=0.97$ and $L=256$ we have performed more than $50$ runs for each 
value of~$p$.
The dynamics are attracted by the empty lattice in all runs 
for $p\le 0.2215$.
On the other hand the active phase is always approached for
$p\ge 0.2219$.
For $p=0.2217$ both phases appear with probability of roughly $1/2$.
Thus an interval exists where both phases are favored by the 
dynamics.
But we observe that this interval decreases with increasing
system size~$L$, i.e., a well defined transition 
point~$p_{\scriptscriptstyle \mathrm{c}}(q)$ exists
in the thermodynamic limit.
For $q=0.97$ the value $p_{\scriptscriptstyle \mathrm{c}}=0.2217$ 
is obtained (see Fig.\,\ref{fig:hysterese_01})
yielding an asymmetric hysteresis between 
$0.221<p<0.223$.

The above analysis of the first-order transition
is performed for a fixed value of~$q$.
Varying~$q$ the first-order transition line
$p_{\scriptscriptstyle \mathrm{c}}(q)$ as well as
the borderlines of supercooling~$p_{\scriptscriptstyle \mathrm{u}}(q)$
and superheating~$p_{\scriptscriptstyle \mathrm{o}}(q)$
can be determined.
In order to limit the numerical effort we focus to the 
determination of $p_{\scriptscriptstyle \mathrm{o}}$.
Since the hysteresis is quite narrow 
($\Delta p/ p_{\scriptscriptstyle \mathrm{c}}\approx 0.009$)
$p_{\scriptscriptstyle \mathrm{u}}$
presents a sufficient approximation of the transition line.
The corresponding values 
are plotted in Fig.\,\ref{fig:phasedia_2d}.
Remarkably, the first-order line ends for $q\to 1$ at
a finite $p$~value in contrast to the mean field phase diagram 
(see Fig.\,\ref{fig:phasedia_mf}).

\section{\textsf{CONCLUSION}}

In summary, we have considered a modification of the 
well established contact process in order to study
the process of tricritical directed percolation.
Taking pair reactions into account the modified
process exhibits a non-trivial phase diagram containing
a tricritical point.
The tricritical point separates a line of second-order
phase transitions from a line of first-order phase transitions.
The transition along the second-order line 
belong to the universality class of directed percolation.
Performing a simple scaling analysis the tricritical point
is determined with high accuracy.
This allows a detailed analysis of the tricritical 
scaling behavior within the steady state.
In particular, we have determined the tricritical 
exponents, universal scaling functions as well as 
universal amplitude ratios.
The obtained values of the critical exponents as well as the
universal amplitude ratios are listed in Table\,\ref{table:exponents}.
Additionally we have investigated the first-order
regime.
A hysteresis is found from steady state measurements.
Owing to effects of metastability supercooling and 
superheating phenomena are observed.
An analysis of the dynamical scaling properties of tricritical
directed percolation will be published elsewhere.\\[1mm]

I would like to thank H.-K.~Janssen, 
A.~Hucht, G.~{\'O}dor, and P.~Grassberger 
for fruitful discussions.


\begin{thebibliography}{10}

\bibitem{HINRICHSEN_1}
H. Hinrichsen, Adv.~Phys. {\bf 49},  815  (2000).

\bibitem{ODOR_1}
G. {\'O}dor, Rev.~Mod.~Phys. {\bf 76},  663  (2004).

\bibitem{LUEB_35}
S. L{\protect\"u}beck, Int.~J.~Mod.~Phys.~B {\bf 18},  3977  (2004).

\bibitem{MARRO_1}
J. Marro and R. Dickman, {\em Nonequilibrium phase transitions in lattice
  models} (Cambridge University Press, Cambridge, 1999).

\bibitem{JANSSEN_4}
{H.-K.~Janssen}, J.~Phys.:~Cond.~Mat. {\bf 17},  S1973  (2005).

\bibitem{JANSSEN_1}
{H.-K.~Janssen}, Z.~Phys.~B {\bf 42},  151  (1981).

\bibitem{GRASSBERGER_2}
P. Grassberger, Z.~Phys.~B {\bf 47},  365  (1982).

\bibitem{LUEB_32}
S. L{\protect\"u}beck and {R.\,D.~Willmann}, Nucl.~Phys.~B {\bf 718},  341
  (2005).

\bibitem{JANSSEN_11}
{H.-K.~Janssen}, Phys.~Rev.~E {\bf 55},  6253  (1997).

\bibitem{JENSEN_10}
I. Jensen, Phys.~Rev.~Lett. {\bf 77},  4988  (1996).

\bibitem{MOREIRA_1}
{A.\,G.~Moreira}, Phys.~Rev.~E {\bf 54},  3090  (1996).

\bibitem{CAFIERO_1}
R. Cafiero, A. Gabrielli, and {M.\,A.~Mu\~{n}oz}, Phys.~Rev.~E {\bf 57},  5060
  (1998).

\bibitem{HOOYBERGHS_1}
J. Hooyberghs, F. Igl{\'o}i, and C. Vanderzande, Phys.~Rev.~E {\bf 69},  066140
   (2004).

\bibitem{VOJTA_2}
T. Vojta and M. Dickison, Phys.~Rev.~E {\bf 72},  036126  (2005).

\bibitem{DOMANY_1}
E. Domany and W. Kinzel, Phys.~Rev.~Lett. {\bf 53},  311  (1984).

\bibitem{ESSAM_1}
{J.\,W.~Essam}, J.~Phys.~A {\bf 22},  4927  (1989).

\bibitem{MANNA_2}
{S.\,S.~Manna}, J.~Phys.~A {\bf 24},  L363  (1991).

\bibitem{ROSSI_1}
M. Rossi, R. Pastor-Satorras, and A. Vespignani, Phys.\,Rev.\,Lett. {\bf 85},
  1803  (2000).

\bibitem{LUEB_26}
S. L{\protect\"u}beck and {P.\,C.~Heger}, Phys.~Rev.~Lett. {\bf 90},  230601
  (2003).

\bibitem{ZHONG_1}
D. Zhong and {D.~ben\,Avraham}, Phys.~Lett.~A {\bf 209},  333  (1995).

\bibitem{CARDY_2}
{J.\,L.~Cardy} and {U.\,C.~T{\protect\"a}uber}, Phys.~Rev.~Lett. {\bf 13},
  4780  (1996).

\bibitem{OHTSUKI_1}
T. Ohtsuki and T. Keyes, Phys.~Rev.~A {\bf 35},  2697  (1987).

\bibitem{OHTSUKI_2}
T. Ohtsuki and T. Keyes, Phys.~Rev.~A {\bf 36},  4434  (1987).

\bibitem{JANSSEN_14}
{H.-K.~Janssen}, {M.~M\"uller}, and O. Stenull, Phys.~Rev.~E {\bf 70},  026114
  (2004).

\bibitem{BASSLER_1}
{K.\,E.~Bassler} and {D.\,A.~Browne}, Phys.~Rev.~Lett. {\bf 77},  4094  (1996).

\bibitem{BAGNOLI_2}
F. Bagnoli, N. Boccara, and R. Rechtman, Phys.~Rev.~E {\bf 63},  046116
  (2001).

\bibitem{ATMAN_1}
{A.\,P.\,F.~Atman}, R. Dickman, and {J.\,G.~Moreira}, Phys.~Rev.~E {\bf 67},
  016107  (2003).

\bibitem{HARRIS_2}
{T.\,E.~Harris}, Ann.~Prob. {\bf 2},  969  (1974).

\bibitem{ODOR_PRIVAT_2005}
G. {\'O}dor, private communication {\bf ~},  ~  (2005).

\bibitem{DICKMAN_18}
R. Dickman and {T.\~Tom{\'e}}, Phys.~Rev.~A {\bf 44},  4833  (1991).

\bibitem{ZIFF_2}
{R.\,M.~Ziff} and {B.\,J.~Brosilow}, Phys.~Rev.~A {\bf 46},  4630  (1992).

\bibitem{BROSILOW_1}
{B.\,J.~Brosilow} and {R.\,M.~Ziff}, Phys.~Rev.~A {\bf 46},  4534  (1992).

\bibitem{LUEB_27}
S. L{\protect\"u}beck and {R.\,D.~Willmann}, J.~Phys.~A {\bf 35},  10205
  (2002).

\bibitem{MORI_1}
H. Mori and {K.\,J.~Mc\,Neil}, Prog.~Theor.~Phys. {\bf 57},  770  (1977).

\bibitem{LUEB_28}
S. L{\protect\"u}beck and {R.\,D.~Willmann}, J.~Stat.~Phys. {\bf 115},  1231
  (2004).

\bibitem{CARDY_1}
{J.\,L.~Cardy} and {R.\,L.~Sugar}, J.~Phys.~A {\bf 13},  L423  (1980).

\bibitem{OBUKHOV_2}
{S.\,P.~Obukhov}, Physica~A {\bf 101},  145  (1980).

\bibitem{LIGGETT_1}
{T.\,M.~Liggett}, {\em Interacting particle systems} (Springer, New York,
  1985).

\bibitem{DICKMAN_7}
R. Dickman, Phys.~Rev.~E {\bf 60},  2441  (1999).

\bibitem{LUIJTEN_1}
E. Luijten, {H.\,W.\,J.~Bl{\protect\"o}te}, and K. Binder, Phys.~Rev.~Lett.
  {\bf 79},  561  (1997).

\bibitem{LUEB_23}
S. L{\protect\"u}beck and {P.\,C.~Heger}, Phys.~Rev.~E {\bf 68},  056102
  (2003).

\bibitem{BINDER_1}
K. Binder and {D.\,W.~Heermann}, {\em Monte Carlo Simulation in Statistical
  Physics} (Springer, Berlin, 1997).

\bibitem{LUEB_33}
S. L{\protect\"u}beck and {H.-K.~Janssen}, Phys.~Rev.~E {\bf 72},  016119
  (2005).

\bibitem{GRASSBERGER_8}
P. Grassberger, J.~Phys.~A {\bf 22},  3673  (1989).

\bibitem{LEE_3}
{J.\,C.~Lee}, J.~Phys.~A {\bf 24},  L377  (1991).

\bibitem{VOIGT_1}
{C.\,A.~Voigt} and {R.\,M.~Ziff}, Phys.~Rev.~E {\bf 56},  R6241  (1997).

\bibitem{GRASSBERGER_3}
P. Grassberger and {Y.-C.~Zhang}, Physica~A {\bf 224},  169  (1996).

\bibitem{JENSEN_6}
I. Jensen, Phys.~Rev.~A {\bf 45},  R563  (1992).

\end{thebibliography}

\end{document}